\documentclass[aps,prx,twocolumn,footinbib]{revtex4-2}
\usepackage{graphicx}
\usepackage{indentfirst}
\usepackage{physics}
\usepackage{braket}
\usepackage{float}
\usepackage{amsmath}
\usepackage{mathtools}
\usepackage{epstopdf}
\usepackage{footnote}
\usepackage{CJK}
\usepackage{esint}
\usepackage{color}
\usepackage[T1]{fontenc}
\usepackage{subfigure}
\usepackage{amsfonts}
\usepackage{footmisc}
\usepackage{scrextend}
\usepackage{multirow}
\usepackage[hyperfootnotes=false]{hyperref}
\usepackage[acronym]{glossaries}
\usepackage{bbm}

\usepackage[english]{babel}
\usepackage{url}
\usepackage{bm}
\usepackage{hyperref}
\definecolor{darkblue}{rgb}{0,0,0.5}
\hypersetup{
colorlinks=true,
linkcolor=black,
filecolor=blue,
citecolor=darkblue,  
urlcolor=black,
}

\newcommand{\calC}{{\cal C}}

\newcommand{\1}{^{(1)}}

\newcommand{\state}[1]{\ketbra{#1}{#1}}
\DeclareMathOperator*{\argmax}{argmax}

\def\be{\begin{equation}}
\def\ee{\end{equation}}
\def\ba{\begin{eqnarray}}
\def\ea{\end{eqnarray}}
\usepackage{bm}

\begin{document}

\title{Fast suppression of classification error in variational quantum circuits}
\author{Bingzhi Zhang$^{1,2}$}
\author{Quntao Zhuang$^{2,3}$}
\email{zhuangquntao@email.arizona.edu}
\affiliation{
$^1$Department of Physics, University of Arizona, Tucson, AZ 85721, USA
\\
$^2$Department of Electrical and Computer Engineering, University of Arizona, Tucson, AZ 85721, USA
\\
$^3$James C. Wyant College of Optical Sciences, University of Arizona, Tucson, AZ 85721, USA
}
\date{\today}

\begin{abstract}
Variational quantum circuits (VQCs) have shown great potential in near-term applications. However, the discriminative power of a VQC, in connection to its circuit architecture and depth, is not understood. 
To unleash the genuine discriminative power of a VQC, we propose a VQC system with the optimal classical post-processing---maximum-likelihood estimation on measuring all VQC output qubits. 
Via extensive numerical simulations, we find that the error of VQC quantum data classification typically decay exponentially with the circuit depth, when the VQC architecture is extensive---the number of gates does not shrink with the circuit depth. This fast error suppression ends at the saturation towards the ultimate Helstrom limit of quantum state discrimination. On the other hand, non-extensive VQCs such as quantum convolutional neural networks are sub-optimal and fail to achieve the Helstrom limit. To achieve the best performance for a given VQC, the optimal classical post-processing is crucial even for a binary classification problem. To simplify VQCs for near-term implementations, we find that utilizing the symmetry of the input properly can improve the performance, while oversimplification can lead to degradation.
\end{abstract}
\maketitle

\section{Introduction}

Quantum computation promises to solve classically intractable problems with a speedup in performance~\cite{Shor_1997}. However, as scalable error-corrected quantum computers are not available, quantum information processing is limited to protocols using noisy intermediate-scale quantum (NISQ)~\cite{Preskill2018quantumcomputingin} technology. The technological constraints also call for an alternative route towards a quantum advantage. Among the candidates, variational quantum circuits (VQCs) are a class of quantum-classical hybrid systems applicable to various tasks, including optimization~\cite{farhi2014quantum}, state preparation~\cite{wecker2015progress,chen2020demonstration}, auto-encoding~\cite{romero2017quantum,gullans2020quantum}, eigen-solvers~\cite{peruzzo2014variational,kandala2017hardware,mcclean2016theory,o2016scalable,colless2018computation,bravo2020scaling, wiersema2020exploring}, unsampling and state approximation~\cite{carolan2020variational,benedetti2019adversarial}, state classification~\cite{patterson2021quantum,chen2021universal,cong2019quantum,maccormack2020branching}, state tomography~\cite{liu2020variational}, sensor networks~\cite{zhuang2019physical,xia2021quantum}, solving partial-differential equations~\cite{lubasch2020variational}, quantum simulation~\cite{li2017efficient,dumitrescu2018cloud,mcardle2019variational} and machine learning in general~\cite{schuld2015introduction,biamonte2017quantum,dunjko2018machine,rebentrost2018quantum,killoran2019continuous,havlivcek2019supervised,schuld2019quantum,du2020expressive,yang2021provable}.

Despite various applications, the fundamental understanding of the capability of VQCs in connection to circuit depth and circuit architecture is still missing. Recent progresses unveil the notion of depth efficiency on expressive and discriminative power in VQCs' classical counterpart---neural networks~\cite{eldan2016power,telgarsky2016benefits,rolnick2017power,lu2017expressive}; VQCs' discriminative power on quantum data has also attracted much attention recently, showing great potential in the classification of few-qubit states~\cite{chen2021universal} and quantum phases of many-body systems~\cite{cong2019quantum,maccormack2020branching}, even in the presence of noise~\cite{patterson2021quantum}.

In this paper, we take a further step to unveil how VQCs' discriminative power quantitatively connects to circuit depth and architecture, and compares with the ultimate limit~\cite{helstrom1976quantum}.
First, to unleash the genuine discriminative power of VQCs, we go beyond the standard approach---a single-qubit measurement on VQC output---to measure all qubits and then perform maximum-likelihood estimation (MLE) on the measurement results, even for a binary classification problem. Our numerical results show that the MLE-VQC approach offers an order-of-magnitude smaller deviation from the Helstrom limit than the single-qubit approach. With the full discriminative power of VQCs in hand, we proceed to explore its connection to circuit architecture and depth.

When discriminating between complex quantum states, we find that the discrimination error is exponentially suppressed with the continuous increase of the VQC depth, until a saturation to the minimum given by the Helstrom limit~\cite{Helstrom_1967,Helstrom_1976}. 
When such a continuous increase of depth is forbidden by the non-extensive architectures, e.g., tree tensor network (TTN), multi-scale entanglement renormalization ansatz (MERA)~\cite{grant2018hierarchical} and quantum convolutional neural networks (QCNNs)~\cite{cong2019quantum}, the discriminative error deviates from the Helstrom limit, even for translation-invariant (TI) or less entangled input states. For extensive architectures that allows the continuous increase of VQC depth, we find that the discriminative power is closely connected to the scrambling power of VQCs. %Overall, the discrimination error is significantly smaller than the approximation error in state generation tasks. 

To reduce the complexity in experimental implementation, we consider simplified VQCs with less parameters or gates. Given the same VQC architecture, for symmetric input states, assuming a symmetric VQC gate parameters makes the VQC much easier to train while still competitive in the error probability performance; for real ground states from many-body systems, restricting the VQC to implement a real unitary significantly reduces the number of gates required to achieve the optimal performance. Indeed, simplification of VQCs helps state discrimination only when properly utilizing the symmetry of the input states.

\section{Circuit architecture and main results}
\label{architecture}

\begin{figure}[t]
    \centering
    \includegraphics[width=0.45\textwidth]{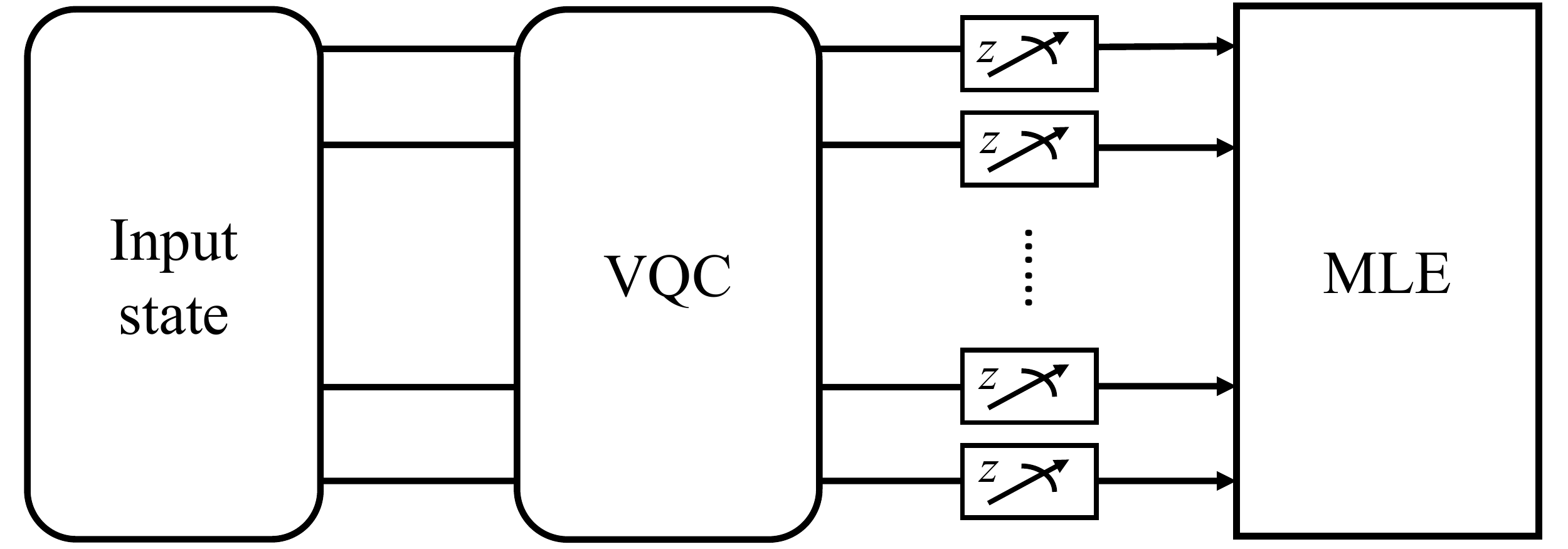}
    \caption{Schematic illustration of the MLE-VQC system. As any single qubit rotation can be combined into the VQC, we can fix each single-qubit measurement to be Pauli $Z$ without loss of generality. An MLE strategy is utilized on the measurement results to make the decision.}
    \label{fig:scheme}
\end{figure}

As shown in Fig.~\ref{fig:scheme}, to perform state discrimination, our MLE-VQC system utilizes a VQC to process the input state, and then performs measurement. Different from existing approaches~\cite{patterson2021quantum,chen2021universal,cong2019quantum,maccormack2020branching}, we consider a measurement on all qubits and optimal MLE post-processing to perform the state discrimination task. In this paper, we will consider multiple-qubit input states, while infinite-dimensional quantum states can be potentially considered generalizing the approach in Ref.~\cite{zhuang2019physical,xia2021quantum}.

A VQC is determined by the type of allowed gates and the circuit architecture. To discriminate between general states, we allow each (two-qubit) gate in the VQC to be universal, composed of single qubit rotations and CNOT gates. As each gate only acts on two qubits, the spread and processing of quantum information is determined by the circuit architecture. We will start with the simple 1-D ``Brickwall'' local circuits (see Fig.~\ref{fig:state_prepare}a or Fig.~\ref{fig:architectures}a) with interchanging between gates acting on two set of neighboring pairs. In Sec.~\ref{sec:benchmarks}, we benchmark between different circuit architectures, including extensive ones (brickwall, prism and polygon~\cite{wu2020scrambling}) and non-extensive ones (QCNN~\cite{cong2019quantum, maccormack2020branching}, TTN and MERA ~\cite{grant2018hierarchical}), as shown in Fig.~\ref{fig:architectures}. We also explore restricted gate sets to simplify VQCs for near-term implementations.

In order to retrieve the maximal information from the states output by the VQC, we perform simultaneous single-qubit measurements on all $n$ qubits in the system, as shown in Fig.~\ref{fig:scheme}. After the measurement, we take the MLE decision strategy, where the decision on the input state is chosen to maximize the posterior probability of the measurement outcome (see Appendix~\ref{app:preliminary} for details). Given the VQC and the final measurements, MLE is known to be the optimal decision strategy that minimizes the error probability.

The minimum `Helstrom' error probability~\cite{Helstrom_1967,Helstrom_1976} further optimizes over the measurement bases, leading to the ultimate error probability of state discrimination (see Appendix~\ref{app:preliminary}). When discriminating between a pair of equal-prior pure states $\{\psi_0,\psi_1\}$, the Helstorm limit has a simple closed-form
\be 
P_{\rm H}\left(\psi_0,\psi_1\right)=\frac{1}{2}\left[1-\sqrt{1-|\braket{\psi_0|\psi_1}|^2}\right].
\label{Helstrom_pure}
\ee 
%In this case, the optimal measuremnt is described by the projective POVM elements onto bases formed by linear superpositions of $\ket{\psi_0}$ and $\ket{\psi_1}$ (see Appendix~\ref{app:preliminary}). 

We train the VQC unitary $U_D$ to achieve the lowest error probability $P_{\rm E}\left(U_D;\psi_0,\psi_1\right)$ in state discrimination (`dis') between states $\{\psi_0,\psi_1\}$. The corresponding cost function for training is chosen to be
\begin{align}
&\calC_{\rm dis}(U_D; \psi_0,\psi_1) \equiv P_{\rm E}\left(U_D;\psi_0,\psi_1\right) - P_{\rm H}\left(\psi_0,\psi_1\right).
\label{eq:cost_dis}
\end{align}
Details for the VQC training process can be found in Appendix~\ref{app:comp_details}. 
Below we summarize the main results in this paper, which are obtained on VQCs after a sufficiently long period of training.
 
In Sec.~\ref{complex_states}, we show that our MLE-VQC strategy exponentially suppresses the error with the growth of the circuit depth until the saturation to the Helstrom limit, when discriminating between complex quantum states. The VQC circuit does so by engineering the output state to be highly entangled such that local measurements can realize complex positive-valued operator measure (POVM) elements. As we show in Sec.~\ref{sec:linear_growth}, when the complexity of input states---quantified by the preparation circuit depth $D_0$---increases, the depth of the VQC circuit required to achieve close to the Helstrom limit increases linearly with $D_0$.

Compared with other tasks such as state generation, we find the VQC state discrimination task is easier, both in terms of the error and trainability, as detailed in Sec.~\ref{sec:comparison}. In addition, symmetry in the inputs makes the Helstrom limit larger, but otherwise preserves all other characters in a VQC state discrimination task.

In Sec.~\ref{app:MLE_vs_one}, we demonstrate the importance of the MLE decision strategy. Given the same circuit architecture and gate set, the performance achievable by a single qubit measurement deviates from the Helstrom limit by an order of magnitude more, compared to the deviation of the MLE case. In addition, the MLE strategy also makes the training easier by increasing the gradients.

Sec.~\ref{sec:architecture} addresses the benchmark between VQC architectures.
Regardless of whether the random input states are symmetric or not, the extensive architectures with a constant number of gates per depth (brickwall, prism, polygon) work much better than those with a limited depth (QCNN, TTN and MERA). This shows a limitation of those over-simplified ansatzs. Due to nonlocal gates, prism and polygon have slight advantages in error probability over the brickwall architecture, consistent with their scrambling powers~\cite{nahum2017quantum,nahum2018operator,zhuang2019scrambling,zhang2021entanglement,wu2020scrambling}, as verified by operator size calculations in Sec.~\ref{scrambling}.

Sec.~\ref{sec:nisq} addresses the simplification of VQCs for near-term implementation.
When the input is symmetric, given the same circuit architecture, the symmetric VQC ansatz works almost as good as the general VQC ansatz, and is much easier to train due to larger gradients. For ground states of a time-reversal symmetric Hamiltonian that have real wave functions, assuming a real matrix representation of the VQC circuit offers similar performance advantages. However, simplifications not based on symmetry and structure of the input can harm the performance.

\section{Complexity of states}
\label{states_complexity}

Quantum state discrimination has a wide range of applications, in quantum communication, quantum sensing and many-body physics, which involves different types of states. In quantum communication, the decoding of classical information can be considered as a quantum state discrimination task, and the direct coding part of capacity theorems are often obtained via a hypothesis testing approach~\cite{datta2013smooth,wang2012one,anshu2017quantum,anshu2018building,hayashi2003general}. There, the state involved can be simple, for example coherent states in optical communication, and can also be entangled across a large number of inputs in more advanced encoding. In quantum sensing, distributed sensing~\cite{zhuang2018distributed,ge2018distributed} and other applications~\cite{tan2008quantum,zhuang2021quantum,Qreading,shi2020entanglement} involve entangled state in a complex form. In many-body physics, people are interested in detecting complex quantum phases of matter, which involves ground states that can be highly entangled~\cite{cong2019quantum,maccormack2020branching}. 
%When the Hamiltonian is gapless, the ground states involved can have `volume-law' entanglement, where the amount of entanglement entropy of a subsystem is proportional to the subsystem volume; While, for gapped Hamiltonians, the ground states have `area-law' entanglement, proportional to the subsystem boundary area.
%%%%in our case we find gapless case is log

\subsection{States generated from local quantum circuits}
\label{sec:state_scrambling}

\begin{figure}[t]
    \centering
    \includegraphics[width=0.475\textwidth]{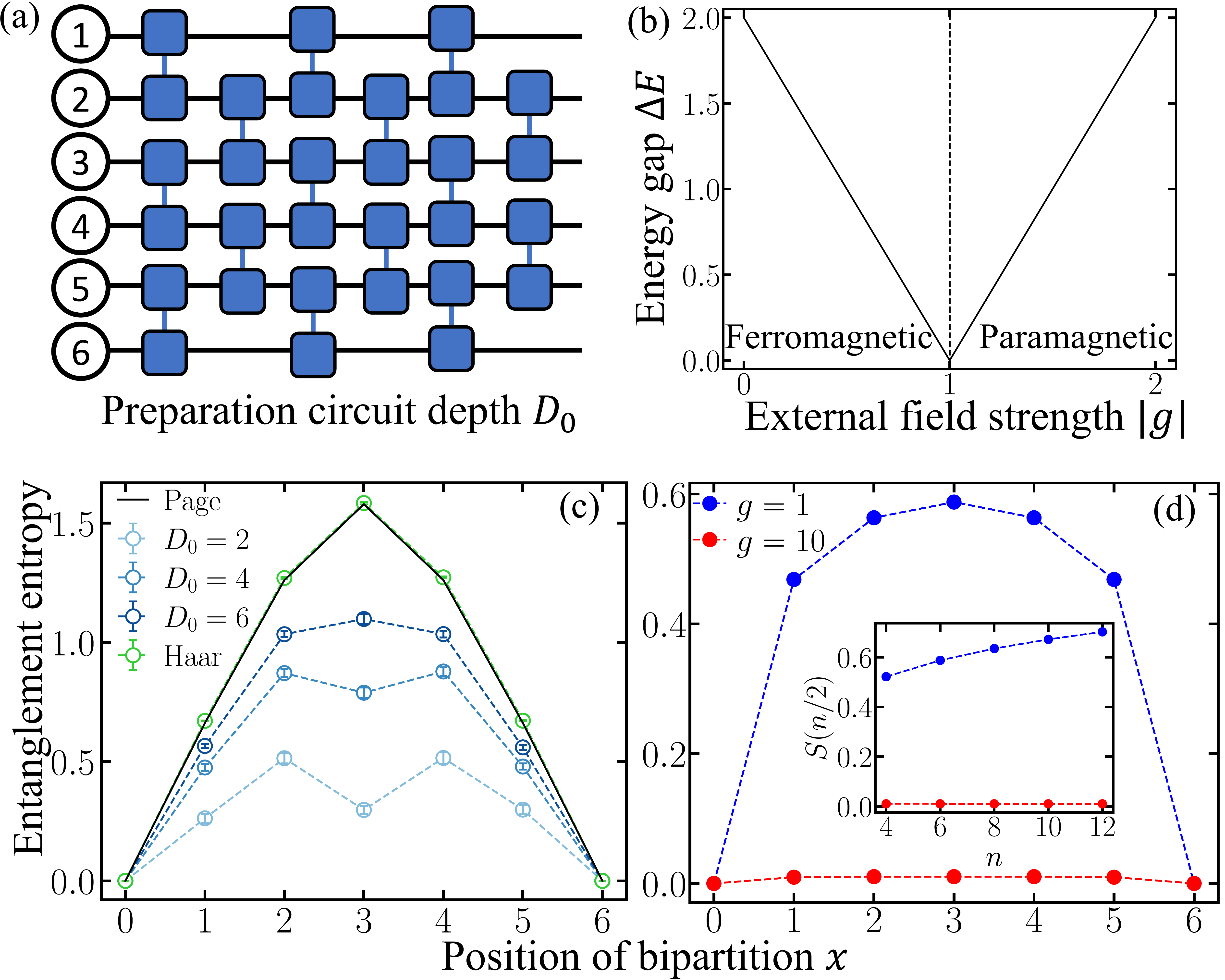}
    \caption{(a) The open boundary local random unitary circuit, ``Brickwall''. Every connected boxes represent local Haar $2-$qubit unitary gate. Here we show an example of a depth $D_0=6$ circuit on $n=6$ qubits input. (b) Phase diagram for TFIM. The black curve represents the analytical expression for energy gap as $\Delta E=2\left(|g|-1\right)$, and the dashed line indicates the critical point. (c)  Ensemble-averaged bipartite von Neumann entanglement entropy of states prepared by circuit in (a). (d) Bipartite entanglement entropy curve for $n=6$ spins TFIM ground states with $g=1, 10$. Inset shows the maximal bipartite entanglement entropy with different $n$. 
    }
    \label{fig:state_prepare}
\end{figure}

To represent different classes of states involved, we consider quantum states generated by inputting a trivial product state $\ket{\bf 0}=\ket{0}^{\otimes n}$ to local quantum circuits, composed of general two-qubit gates acting on neighboring qubits (see Fig.~\ref{fig:state_prepare}(a)). As we choose the gates randomly, the ensemble is characterized by the preparation circuit depth $D_0$, and therefore denoted as $\mathbb{H}\left({D_0}\right)$. The circuit complexity and entanglement of typical states in $\mathbb{H}\left({D_0}\right)$ grow linearly with the depth $D_0$~\cite{eisert2021entangling,haferkamp2021linear} before saturation. 
As shown in Fig.~\ref{fig:state_prepare}(c), when $D_0<n$ is a fixed constant, the states generated have area-law entanglement; while when $D_0\propto n$ is large, the states is typically highly entangled under a volume-law. 
Indeed, $\mathbb{H}\left({D_0}\right)$ well captures the different problems of interest. In quantum communication, states in $\mathbb{H}\left({D_0}\right)$ are used as the random encoding~\cite{gullans2020quantum} to achieve capacity; in many-body physics, the depth $D_0$ will control the bound dimension of the matrix product representation of states. Moreover, $\mathbb{H}\left({D_0}\right)$ are also studied in quantum information scrambling~\cite{nahum2017quantum,nahum2018operator} and t-design complexity~\cite{gross2007evenly,ambainis2007quantum,roberts2017chaos,brandao2016local}.

%random circuits are considered to represent chaotic dynamics of many-body systems~\cite{nahum2017quantum,nahum2018operator}. Operators spread at a so-called butterfly speed $v_B$, while entanglement grows linearly $\sim v_E D$, with a speed $v_E<v_B$. The random choice also connects to the t-design complexity measure of the unitary~\cite{gross2007evenly,ambainis2007quantum,roberts2017chaos}. As local random circuits with depth linear in the system size $n$ can approximate $t$-design~\cite{brandao2016local}, we expect the `complexity' of the state discrimination problem to grow with the depth $D_0$ of generation circuits.

For $D_0\to\infty$, the ensemble $\mathbb{H}\left(\infty\right)$ approaches Haar random states and the typical Helstrom limit between states $\psi_0,\psi_1\in \mathbb{H}\left(\infty\right)$ can be evaluated
\be 
\expval{P_{\rm H}\left(\psi_0,\psi_1\right)}_{\mathbb{H}\left(\infty\right)}=\frac{1}{2\left(2^{n+1}-1\right)}\sim \frac{1}{2^{n+2}},
\label{PH_D0gg1}
\ee 
For a finite $D_0$, when $2^{n+1}\gg1$, Eq.~\eqref{PH_D0gg1} still holds to the leading order (see Appendix~\ref{app:haar} for details).

As many-body systems often have a translational symmetry. We therefore also consider the subset of TI states $\mathbb{S}\left(D_0\right)$, which is prepared by the periodic boundary TI local random unitary circuit with depth $D_0$. The typical Helstrom limit for $\mathbb{S}\left(D_0\right)$ is larger than Eq.~\eqref{PH_D0gg1} for $\mathbb{H}\left(D_0\right)$, however still independent of $D_0$ to the leading order (see Fig.~\ref{fig:mean_helstrom}).

\subsection{Ground states of many-body systems}

We also consider ground states of many body systems.
We focus on the well-known toy model, transverse-field Ising model (TFIM), whose hamiltonian is
\begin{equation}
    H_{\rm TFIM} = -\sum_i Z_i Z_{i+1} + g\sum_i X_i,
\end{equation}
where $Z_i, X_i$ are Pauli matrices at site $i$ and $g$ is the strength of external field relative to the coupling strength. To reduce the finite-size effects, we consider a periodic boundary condition. As depicted in Fig.~\ref{fig:state_prepare}(c), when $|g|<1$, the system stays in an ordered ferromagnetic phase; as $|g|$ increases, it transits to disordered paramagnetic phase. In both phases, the ground states of the system show area-law entanglement (Fig.~\ref{fig:state_prepare}(d)). At the critical point, $|g|=1$, the system undergoes a quantum phase transition. The entanglement entropy shows a logarithmic scaling behavior, which can also be described by conformal field theory with central charge $c=1/2$~\cite{holzhey1994geometric,latorre2003ground,calabrese2004entanglement}.

%In this paper, we consider two different ensembles, random states prepared by deep depth local Haar random unitary circuits, which are expected to be highly entangled, and ground states of many-body systems, which can be described accurately by MPS representation with low entanglement and hold specific symmetries. Optimized VQCs are applied to the two ensembles separately to perform binary state discrimination within each ensemble.

\section{Performance of the brickwall VQC in state discrimination}
\label{sec:local_VQC}

\begin{figure}[t]
    \centering
    \includegraphics[width=0.475\textwidth]{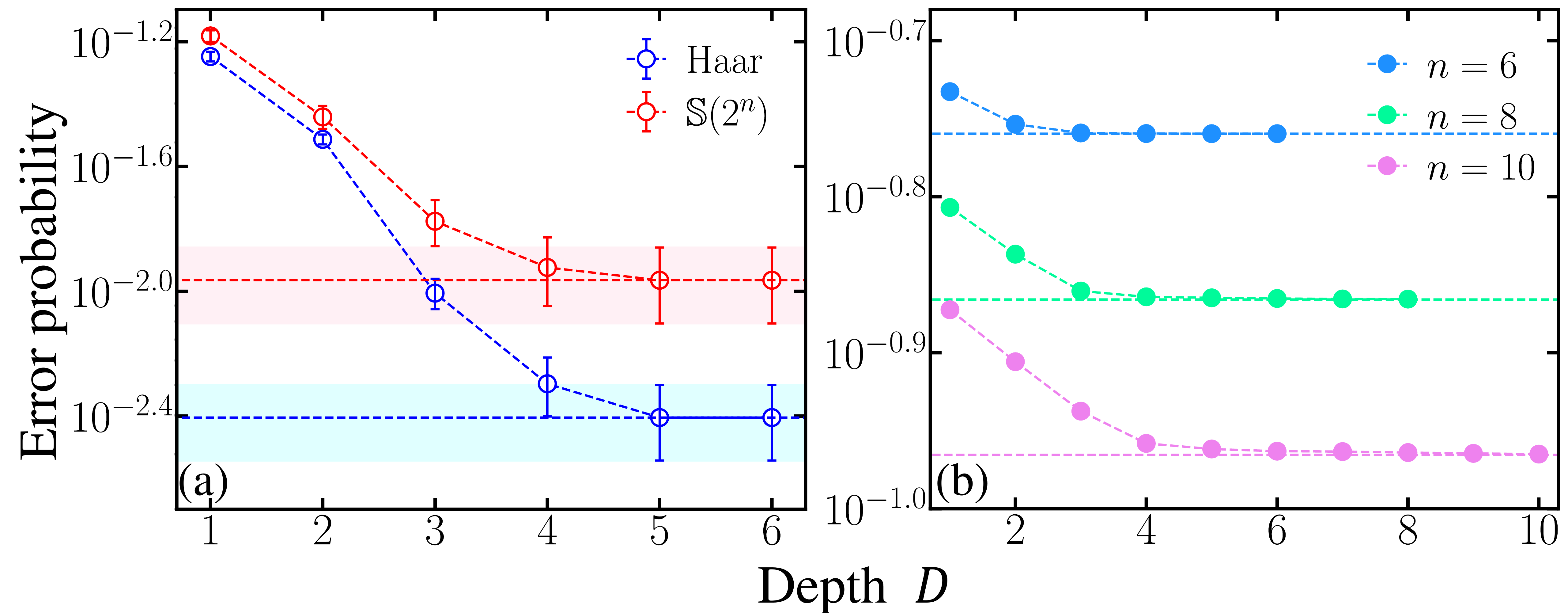}
    \caption{Binary state discrimination error probability with trained MLE-VQC systems. (a) State discrimination between Haar random states or complex TI states $\mathbb{S}\left(2^n\right)$ in a system of $n=6$ qubits. Horizontal dashed lines represent the ensemble-averaged Helstrom limit $\braket{P_{\rm H}}$ and light color area represents the amount of ensemble fluctuation. (b) State discrimination between ground states of TFIM with $g=1$ and $g=10$ in a system of $n=6, 8, 10$ qubits. Horizontal dashed lines show the Helstrom limit in each case.  We use the open-boundary brickwall VQC ansatz in all cases.
    }
    \label{fig:SCob_pe}
\end{figure}

With the VQC circuit depth increasing, the performance of the MLE-VQC system will eventually approach the ultimate Helstrom limit for pure state discrimination. This is because for an ensemble of pure states being considered in this paper, the optimal POVM elements are also rank-one projectors~\cite{kennedy1973optimum,eldar2003designing}; therefore additional ancilla is not necessary in the measurement. 
At the same time, however, the training of the circuit becomes harder as the number of parameters increases and the gradient decays~\cite{mcclean2018barren,cerezo2021cost,wang2020noise,arrasmith2020effect,cerezo2021higher,arrasmith2021equivalence,holmes2021barren}. 

In this section, we explore the error probability performance and trainability of the MLE-VQC system with the open-boundary brickwall VQC ansatz. The periodic-boundary VQC shows similar results, as we shown in Appendix~\ref{app:boundary}. To understand the performance with a finite depth circuit, we evaluate the decay of error probability towards the Helstrom limit and the decay of gradient with depth, for different ensemble of input states.

\subsection{Fast error suppression}
\label{complex_states}

To begin with, we consider the average error probability for complex states discriminated by VQCs with different depth $D$. In Fig.~\ref{fig:SCob_pe}(a), we consider Haar random states (blue dots) and find a fast error suppression--the error probability decays exponentially with $D$ before saturation to the Helstrom limit (blue dashed horizontal line). To represent the symmetric case, we further consider the set of states $\mathbb{S}(2^n)$ prepared by TI local quantum circuits with a large enough depth $2^n\gg1$. While symmetry increases the Helstrom limit (red dashed horizontal line), it does not change the exponential suppression of error probability with depth $D$(red dots). To extend the results beyond random states, we consider the discrimination between two ground states of TFIM with different parameters $g$ in Fig.~\ref{fig:SCob_pe}(b). We see as the number of qubits $n$ increases, the Helstrom limit decreases and the error probability shows an exponentially suppressed trend with VQC depth $D$. Although the number of qubits is limited due to the increasing level of difficulty in the training, we see the depth required to saturate the Helstrom limit scales linearly with the system size.

It is worthy to point out that the amount of entanglement in states before the final measurement is high in the MLE-VQC approach. For the Haar random input states, the optimal VQC of different depth $D$ preserves the bipartite entanglement entropy $\braket{S(n/2)}$ at the Page curve value, as shown by the purple line in Fig.~\ref{fig:pe_dc}(b). Note that for less entangled inputs $\mathbb{H}(D_0)$ prepared by random local circuits of depth $D_0$, the VQC circuit increases the level of entanglement $\braket{S(n/2)}$ at the output side before the final measurements (green, red, blue for $D_0=2,4,6$). From this, we see that the VQC is essentially sorting and increasing entanglement between the qubits to enable the best performance on the final separable measurement.

\subsection{Linear growth growth of complexity}
\label{sec:linear_growth}

\begin{figure*}
    \centering
    \includegraphics[width=0.7\textwidth]{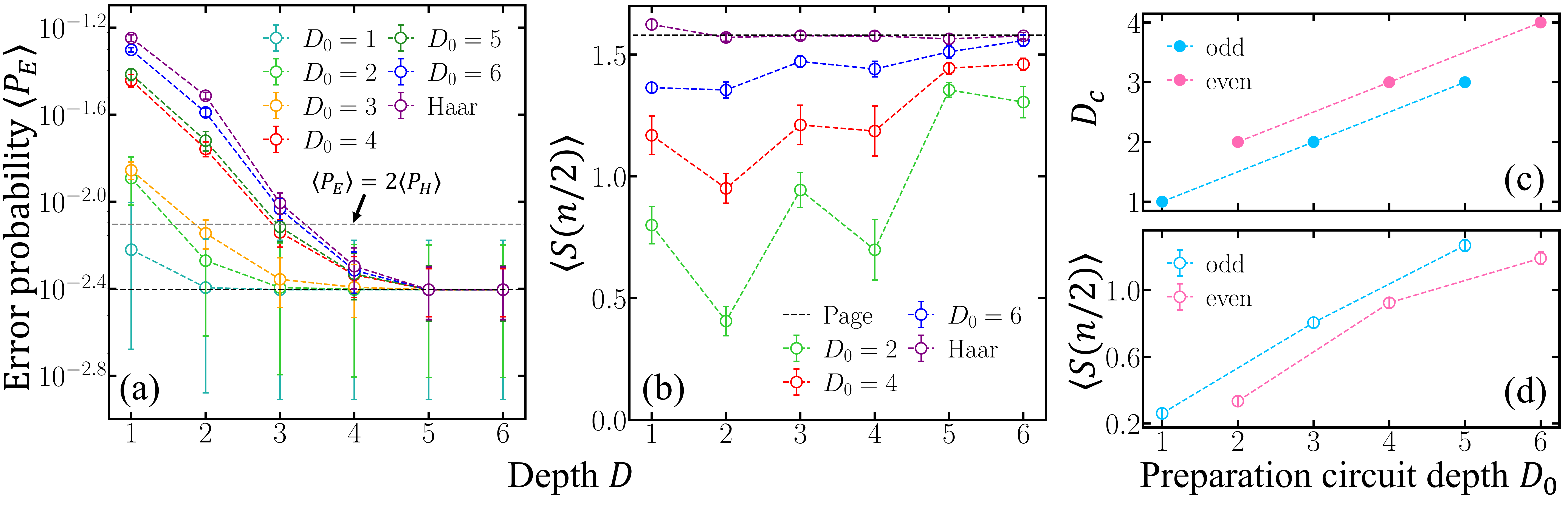}
    \caption{(a) Discrimination error probability using the brickwall ansatz between a pair of random states sampled from $\mathbb{H}(D_0)$ for $n=6$ qubits. Black and grey dashed horizontal lines show the Helstrom limit $\braket{P_{\rm H}}$ and a critical value $\braket{P_E} = 2\braket{P_{\rm H}}$. Note the relative large error bar in the error probability in states $\mathbb{H}(D_0)$ generated by a shallow depth VQC is due to the lack of self-averaging. (b) Average maximal bipartite entanglement entropy $\braket{S(n/2)}$ of VQC output states before measurements for $\mathbb{H}(D_0)$ discrimination. (c) The critical depth $D_c$ to achieve $\braket{P_E}=2\braket{P_H}$ versus input complexity $D_0$. (d) Maximal bipartite entanglement entropy $\braket{S(n/2)}$ of the input states to be distinguished. In (c) and (d), we plot odd and even $D_0$ separately.}
    \label{fig:pe_dc}
\end{figure*}

Here we explore how the complexity of the input state ensemble affects the error probability. As explained in Section~\ref{sec:state_scrambling}, we can tune the complexity of output states $\mathbb{H}\left({D_0}\right)$ produced in a depth-$D_0$ local random circuit  by controlling the depth $D_0$; therefore, we study the discrimination between states sampled from $\mathbb{H}\left({D_0}\right)$.

For states sampled from $\mathbb{H}(D_0)$, the fast suppression of error probability still holds, as shown in Fig.~\ref{fig:pe_dc}(a). With increasing input states complexity, the discrimination task becomes harder, leading to an increasing error probability for a fixed VQC depth $D$. Similar to results in Fig.~\ref{fig:pe_dc}(a), as for binary discrimination between TI states sampled from $\mathbb{S}(D_0)$, the universal exponential suppression still holds with a slightly increased Helstrom limit, as shown in Fig.~\ref{fig:symmetries}.

As the saturation towards the Helstrom limit has a long tail, we consider the number of layers $D_c$ required to achieve an error probability $P_E=2P_{\rm H}$. In Fig.~\ref{fig:pe_dc}(c), we see a linear growth of $D_c$ with $D_0$, as expected. This can also be explained by a sub-optimal strategy mimicking the Kennedy receiver~\cite{kennedy1973receiver}, which implements the POVM element $\Pi_0=\state{\psi_0}$ with a depth $D\sim D_0$ VQC to achieve the error probability $|\braket{\psi_0|\psi_1}|^2/2\simeq 2 P_{\rm H}$ when $P_{\rm H}\ll1$. One can also understand the increase from the increase of entanglement in the input state, which also shows a linear trend, as depicted in Fig.~\ref{fig:pe_dc}(d). Our error probability results are obtained from a finite system of six qubits, however, extending to larger systems to further consolidate the conclusion is challenging, due to the exponential decay of gradient shown in Sec.~\ref{sec:comparison}.

\subsection{Comparison between state generation and discrimination: performance and trainability}
\label{sec:comparison}

To understand the level of difficulty of state discrimination (`dis'), we benchmark with the most relevant task of state generation (`gen')~\cite{carolan2020variational,benedetti2019adversarial}. In an $n$-qubit state generation task, the VQC performs a unitary $U_D$ on a trivial product state $\ket{\bf 0} = \ket{0}^{\otimes n}$ to approximate a target state $\ket{\psi}$. Similar to the discrimination case in Eq.~\eqref{eq:cost_dis}, we utilize the following cost function
\begin{align}
&\calC_{\rm gen}(U_D;\psi) \equiv 1-|\braket{\psi|U_D|\bf 0}|^2, \label{eq:cost_gen} 
\end{align}
as a function of the VQC unitary $U_D$; Details for the numerical optimization process can be found in Appendix~\ref{app:comp_details}.

From Fig.~\ref{fig:SC_gap}(a), we can identify a sharp transition in the cost function $\calC$ for both discrimination and generation, where $\calC$ is exponentially suppressed before reaching an extremely small value. In addition to the overall trend, we can find that $\calC_{\rm dis}$ is about an order of magnitude smaller than $\calC_{\rm gen}$. The gap indicates that for a given depth VQC ansatz, the performance for discrimination is better than generation. The same conclusion is also confirmed for random states sampled from $\mathbb{H}(D_0)$ and $\mathbb{S}(D_0)$ with a finite $D_0$ (see Fig.~\ref{fig:SCob_archs}). Although we have focused on the brickwall ansatz in Fig.~\ref{fig:SC_gap}(a) in this section, the same gap between discrimination and generation exists in other architectures as we will discuss in Sec.~\ref{sec:benchmarks}.

To explore the trainability of VQCs, we evaluate the gradient of the cost functions to the parameters. For both the generation and discrimination tasks, the cost function gradient $g_i$ with respect to parameter $\theta_i$ can be obtained numerically from a central finite-difference. Note that as the gradient can be positive or negative, we evaluate the variance ${\rm Var}(g_i)$ among different positions to get a sense of the magnitude of gradients, similar to Ref.~\cite{mcclean2018barren}. Moreover, we take an average over the different gradient directions to obtain the average variance of gradient $\braket{{\rm Var}(g_i)}_i$.
In Fig.~\ref{fig:SC_gap}(b), the parameter-averaged variance of gradient decays exponentially with the number of qubits $n$, predicted by the well-known barren plateau phenomena~\cite{mcclean2018barren}. We find that the gradient decay to be much faster in the generation case, which indicates that in terms of trainability, discrimination is much easier than generation for states with/without translational symmetry.

\subsection{MLE's superiority over single measurement schemes}
\label{app:MLE_vs_one}

\begin{figure}[t]
    \centering
    \includegraphics[width=0.475\textwidth]{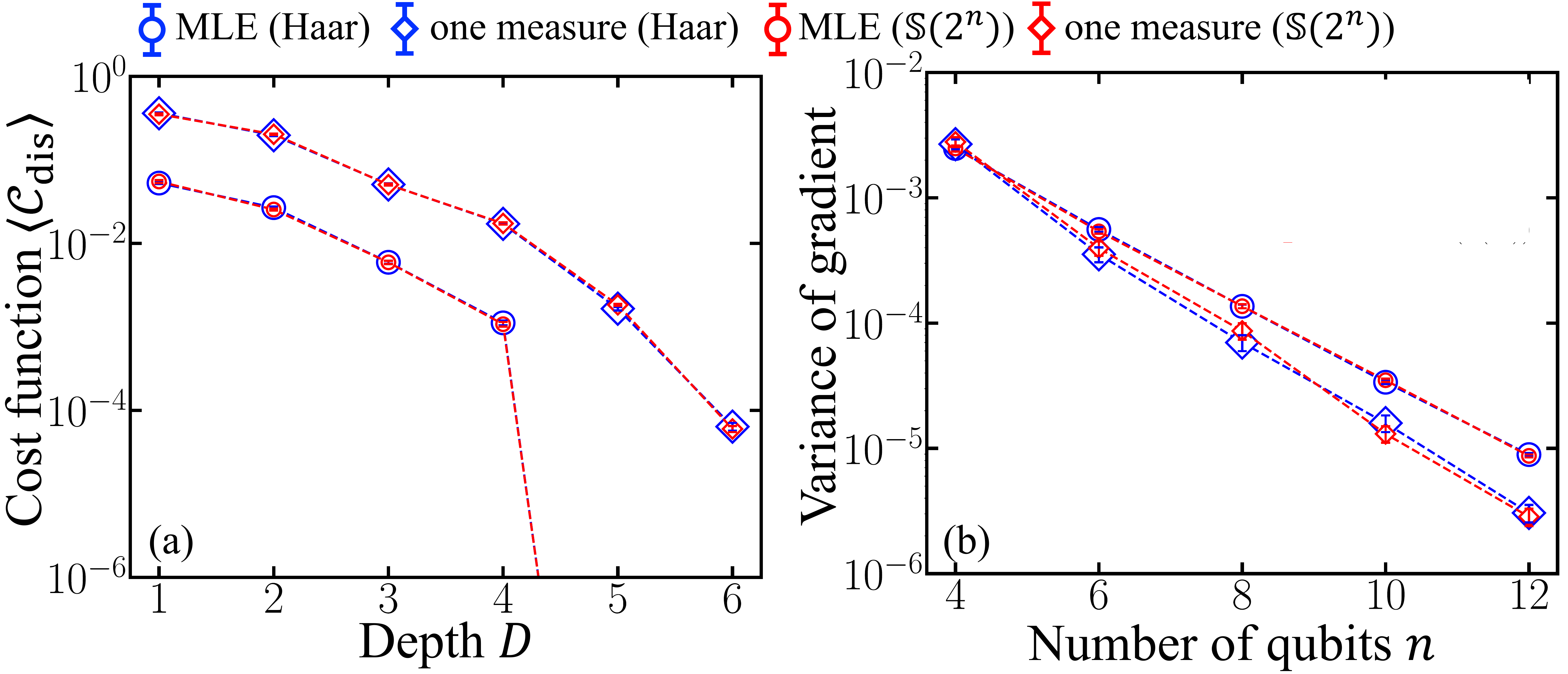}
    \caption{MLE versus single-qubit measurement approach in a brickwall ansatz VQC for discriminating between Haar random states and TI states sampled from $\mathbb{S}(2^n)$. We consider a system of $n=6$ qubits. We show the ensemble-averaged cost function $\braket{\calC_{\rm dis}}$ in (a) and the parameter-averaged variance of gradient $\braket{{\rm Var}\left(g_i\right)}_i$ for ansatz with depth $D=2$ in (b).}
    \label{fig:MLE_vs_one}
\end{figure}

The simultaneous single-qubit measurements on all qubits and optimal MLE decision rule are crucial for our MLE-VQC approach to unleash the full power of the brickwall ansatz. To demonstrate such, we benchmark the MLE-VQC approach against the VQC with only a single-qubit $Z-$measurement at the center. To show their difference, we still focus on two ensembles, Haar random states and complex TI states $\mathbb{S}(2^n)$. As shown in Fig.~\ref{fig:MLE_vs_one}(a), the residual error utilizing MLE-VQC approach is around an order of magnitude smaller than that of the single-qubit approach at a given $D$, which shows the power of MLE to gather the full information from all qubits. Moreover, the MLE case shows a sharp drop in the error at a large enough VQC depth $D$, while the single-qubit case has a consistent decay. As for trainability, although the variance of gradient for both approaches decreases exponentially with the number of qubits $n$, the MLE approach typically has a larger gradient and therefore is easier to train (see Fig.~\ref{fig:MLE_vs_one}(b)).

\section{Performance benchmarks of different ansatz} 
\label{sec:benchmarks}
In Sec.~\ref{sec:local_VQC}, we employ the brickwall VQC ansatz, where each two-qubit gate is applied to pairs of nearest neighbors in one dimension. As we mentioned in Sec.~\ref{architecture}, various other architectures have been proposed for different tasks. Since we find that VQCs relying on increasing the amount of entanglement in the quantum states to approach the Helstrom limit, we expect that architectures with non-local gates might improve the discriminative power of VQCs. In Sec.~\ref{sec:architecture}, we offer a benchmark between different architectures to confirm this. 

\begin{figure}[t]
    \centering
    \includegraphics[width=0.475\textwidth]{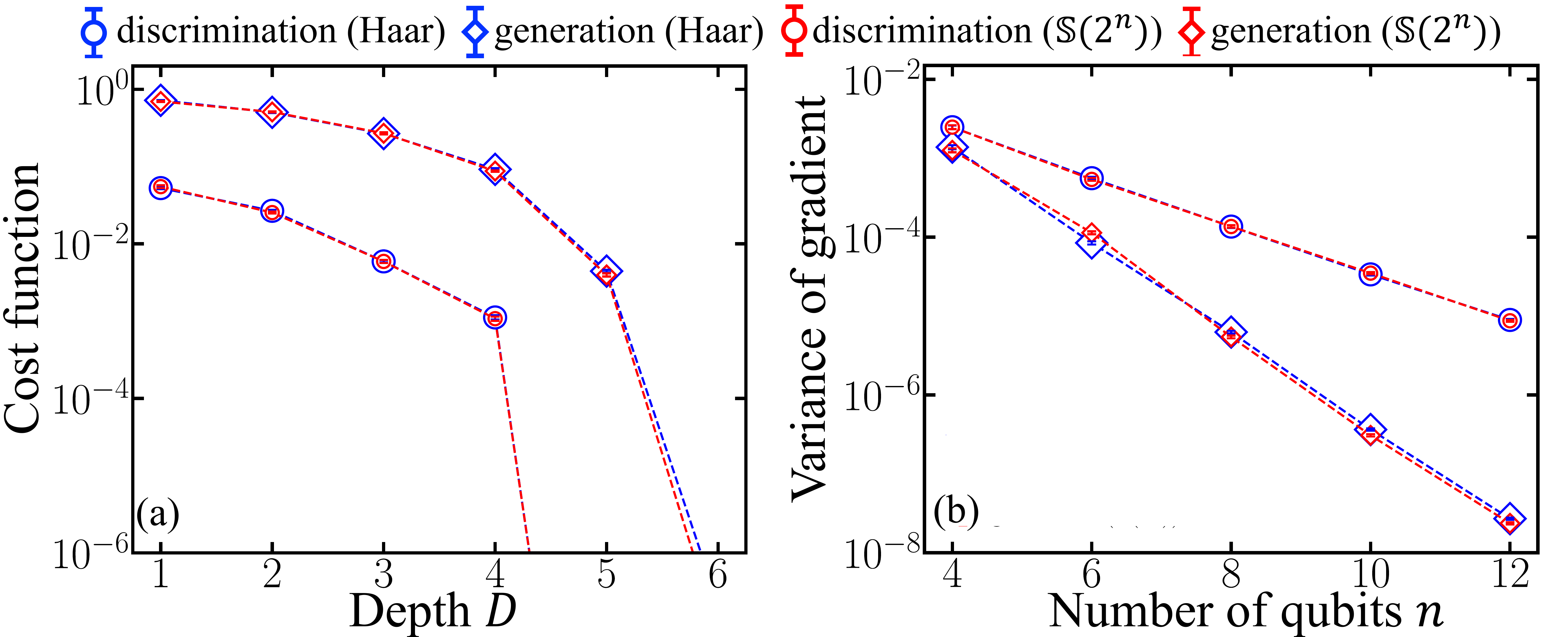}
    \caption{Performance and gradient of the open-boundary brickwall VQC. We consider two different state ensemble, Haar random states and TI states $\mathbb{S}(2^n)$. (a) Ensemble-averaged cost functions for discrimination and generation. Blue and red curves show ensemble of Haar random states and TI states separately in a system of $n=6$ qubits. (b) Parameter-averaged variance of gradient $\braket{{\rm Var}\left(g_i\right)}_i$ in the $D=2$ ansatz.}
    \label{fig:SC_gap}
\end{figure}

Despite the different architectures, in the NISQ era, VQC implementations are limited in the circuit depth and number of gates, due to the accumulation of device imperfections. Therefore, we further explore simplification of the gate sets in Sec.~\ref{sec:nisq}. In particular, we find that symmetry in the input states allows VQCs to be simplified.

\subsection{Comparison between different architectures} 
\label{sec:architecture}

\begin{figure}[t]
    \centering
    \includegraphics[width=0.45\textwidth]{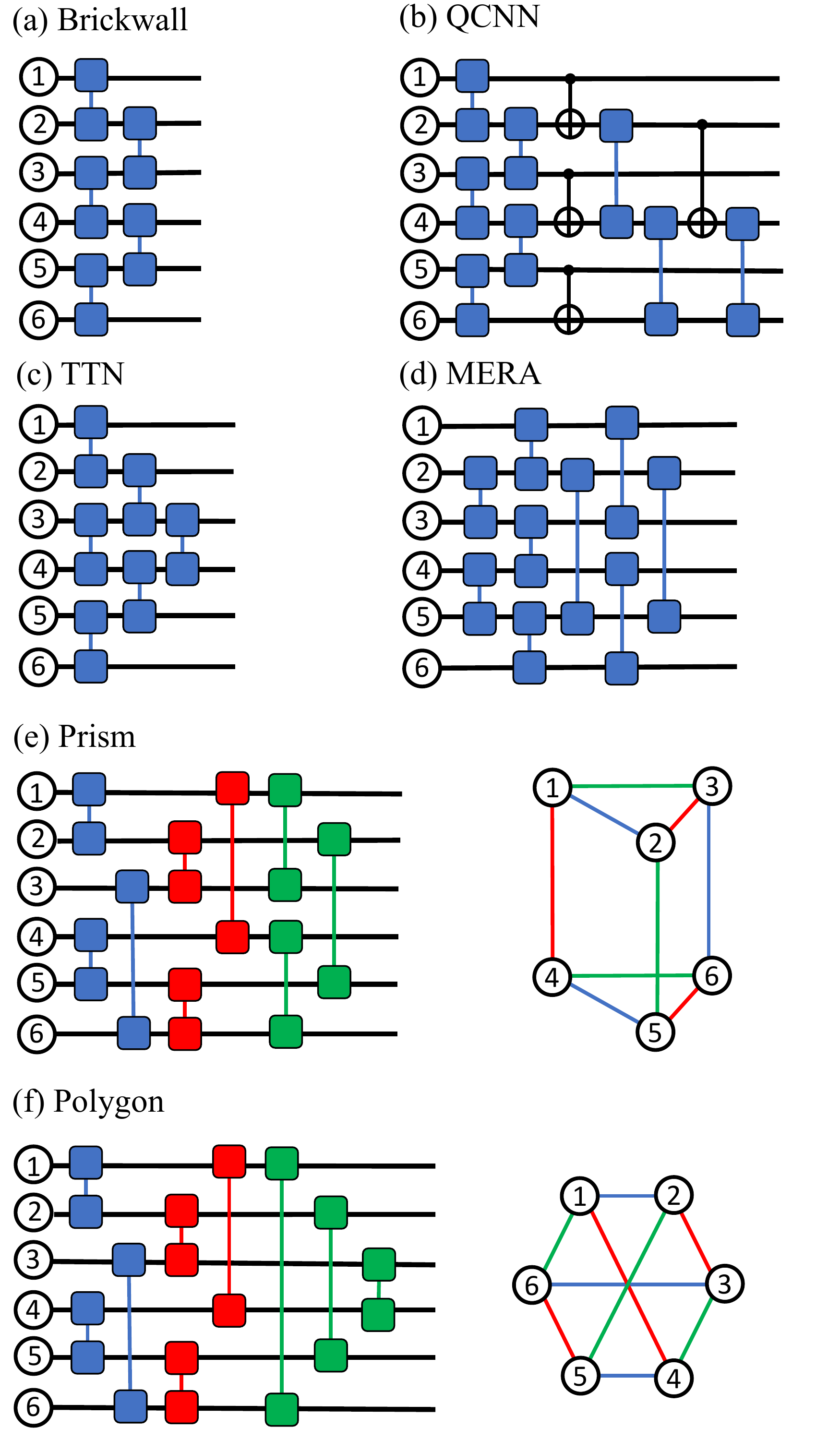}
    \caption{Topological architectures of VQCs. Two connected boxes represents a universal $2-$qubit gate. For (a) brickwall, (e) prism and (f) polygon ansatzs, we show the case with depth $D=2, 3, 3$ separately.}
    \label{fig:architectures}
\end{figure}

In this section, we benchmark the various VQC architectures (see Fig.~\ref{fig:architectures}) for both state discrimination and generation tasks. An architecture determines the layout of the quantum gates and therefore constrain the information flow in the VQC. In Sec.~\ref{sec:local_VQC}, we have focused on the two-local brickwall ansatz. To extend the interactions beyond two-local, prism and polygon~\cite{wu2020scrambling} architectures generalize the line geometry to different shapes. These three architectures are extensive---they have the number of gates per layer roughly unchanged as the depth of the circuit increases. Other popular choices of architectures have a fixed depth and therefore are not extensive, including QCNN~\cite{cong2019quantum, maccormack2020branching} and tensor network architectures (TTN and MERA)~\cite{grant2018hierarchical}.

\begin{figure}[t]
    \centering
    \includegraphics[width=0.475\textwidth]{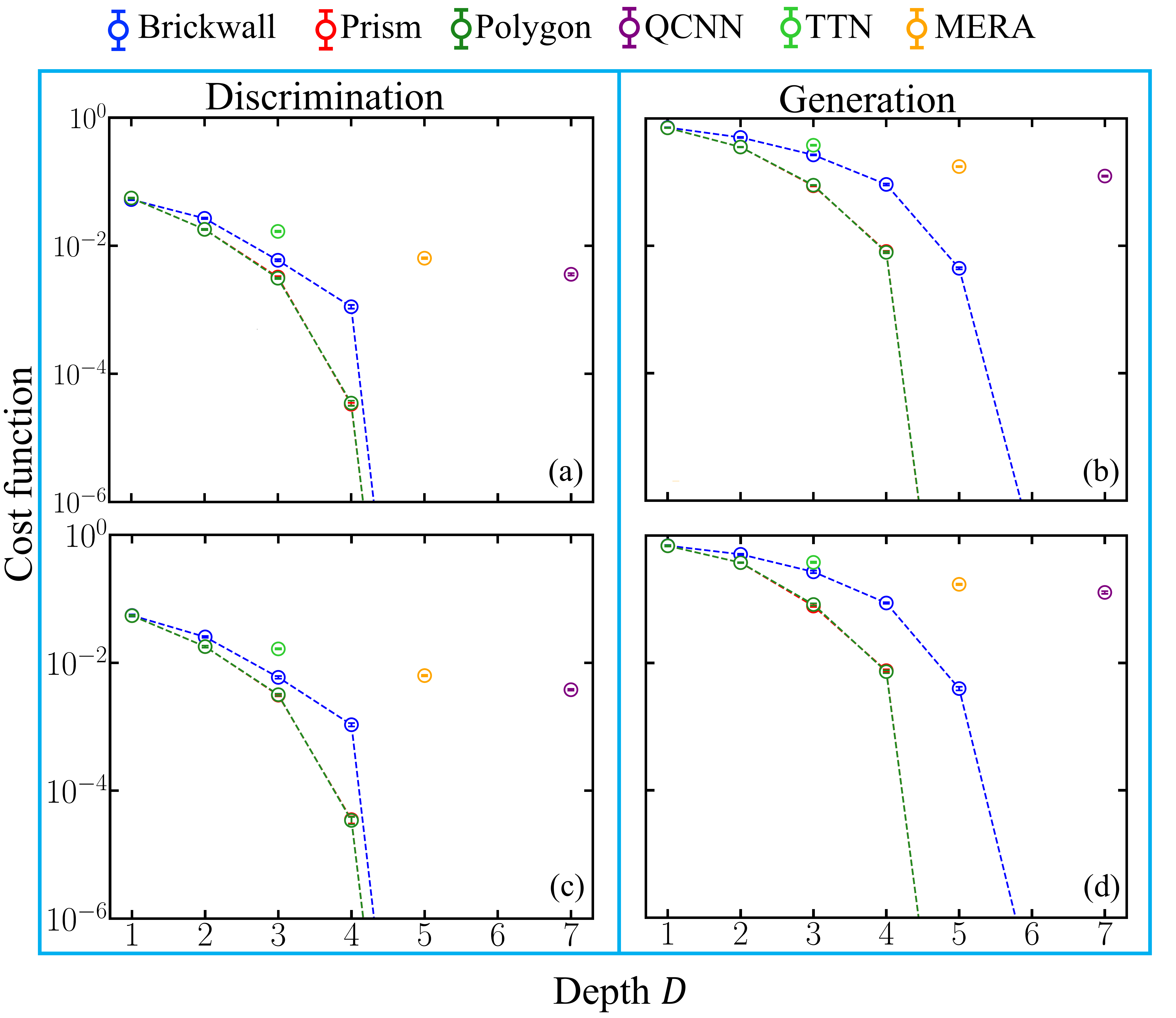}
    \caption{Cost functions for the discrimination (left) and generation (right) among Haar random states (top) or complex TI states $\mathbb{S}(2^n)$ (bottom) in a system of $n=6$ qubits. We compare performance with different VQC architectures. For comparison, the Helstrom limit $\braket{P_H}\sim10^{-2.4}$ for (a) and $\braket{P_H}\sim10^{-2}$ for (c), as shown in Fig.~\ref{fig:SCob_pe}. The relative differences in the cost functions between top and bottom panels are below 3\%.}
    \label{fig:Haar_architectures}
\end{figure}

In Fig.~\ref{fig:Haar_architectures}(a), we begin the benchmark with the error probability performance in discriminating Haar random state pairs. We see that the extensive architectures (brickwall, prism and polygon) provide a better performance over the non-extensive architectures (QCNN, TTN and MERA) at the same depth $D$. In particular, the extensive ones saturate the Helstrom limit at $D=5$ exactly, while the non-extensive ones are far from the Helstrom limit even at a larger depth. Among the three extensive architectures, we find prism and polygon to be slightly better than the brickwall architecture at a finite depth $D$, due to non-local gates more efficiently processing the global quantum information. The same conclusions also generalize to the complex TI inputs $\mathbb{S}(2^n)$, as shown in Fig.~\ref{fig:Haar_architectures}(c). 

We also consider the task of state generation, and find similar conclusions to hold---the relative ordering of the error is identical to that in state discrimination, as shown in Fig.~\ref{fig:Haar_architectures}(b)(d) for the Haar ensemble and TI ensemble. 
This shows a consistent ordering of quantum information processing power among the architectures, which also agrees with the quantum information scrambling capabilities explored in Sec.~\ref{scrambling}. Comparing between the state generation task in Fig.~\ref{fig:Haar_architectures}(b)(d) and the state discrimination task in Fig.~\ref{fig:Haar_architectures}(a)(c), we also extend the previous conclusion in Sec.~\ref{sec:comparison} that state discrimination is easier than state generation to VQCs with different architectures.

Comparing Fig.~\ref{fig:architectures} bottom panels (c,d) with the top panels (a,b), we can see that for a general VQC, translational symmetry in the input merely increases the Helstrom limit, while at any depth the deviations to the Helstrom limit (the cost function) are almost identical with and without input symmetry. Although here symmetry does not make much difference, it will allow simplifications of the VQC, as we will explain in Sec.~\ref{sec:nisq}.

Finally, we want to emphasize that even for less complex states $\mathbb{H}(D_0)$ and $\mathbb{S}(D_0)$ prepared by low-depth circuits with and without symmetry, the non-extensive architectures (QCNN, TTN and MERA) are still worse than the extensive architectures at the same depth, as shown in Fig.~\ref{fig:SCob_archs} and Fig.~\ref{fig:tsSC_archs} in Appendix~\ref{app:other_performance}. In this sense, these non-extensive architectures trade-off performance for the smaller number of quantum gates and parameters. To obtain the optimal performance, extensive architectures are preferable.

\subsection{Discriminative power versus scrambling power}
\label{scrambling}

As the circuit depth increases, a VQC generates more entangled outputs in order to achieve a better error probability in state discrimination (see Fig.~\ref{fig:pe_dc}(b)). Because entanglement growth is also an important indicator of quantum information scrambling in the circuit, we evaluate the scrambling power of the VQCs utilized in Sec.~\ref{sec:benchmarks}, in comparison to their performances. 

\begin{figure}[t]
    \centering
    \includegraphics[width=0.3\textwidth]{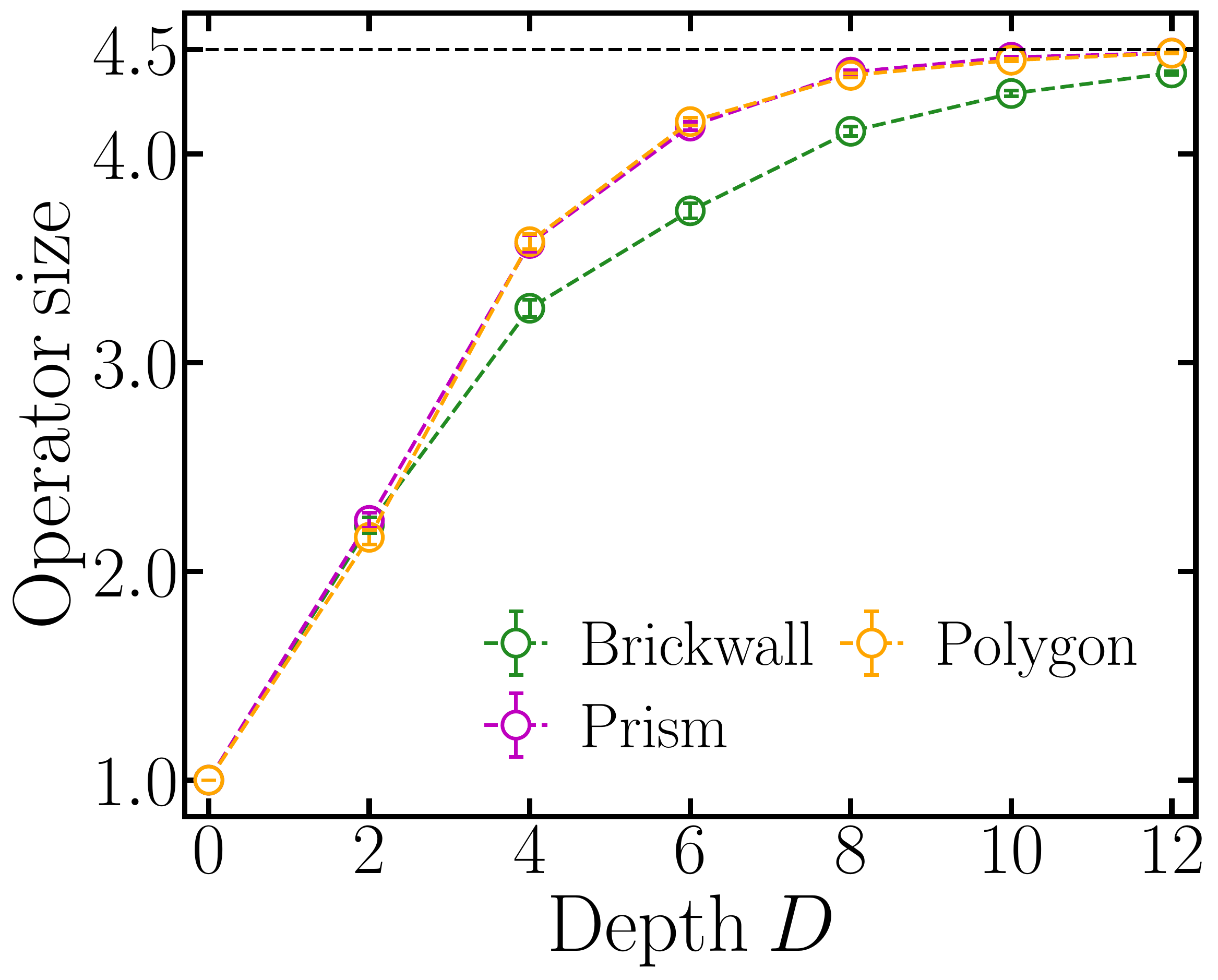}
    \caption{Ensemble-averaged operator size growth with the VQC depth for $Z$ initially localized at the $n/2$-th qubit in a $n=6$ system.}
    \label{fig:operator_size}
\end{figure}

Similar to Ref.~\cite{wu2020scrambling}, we choose the operator size~\cite{zhuang2019scrambling} as the metric to evaluate the scrambling power of VQCs. To define operator size, we consider a Pauli-Z operator $M_0 = I_1\otimes\cdots Z_{n/2}\otimes\cdots\otimes I_n$ initially located at the center, where $I_k$ is an identity operator acting on the $k$th qubit. Under the VQC represented by unitary $U_D$, the operator evolves to $M_D=U_D^\dagger M_0 U_D$. In general, $M_D$ can be expanded in the Pauli bases, i.e., $M_D=\sum_\mathcal{S} \alpha_{\mathcal{S}} \mathcal{S}$, where $\mathcal{S}=\otimes_{k=1}^n \sigma_k$ is a Pauli string with $\sigma_k$ being one of the four Pauli operator $I_k, X_k, Y_k, Z_k$ at $k$th qubit. The size of the evolved operator
\be
{\rm Size}\left(M_D\right) = \sum_\mathcal{S} |\alpha_{\mathcal{S}}|^2 L\left(\mathcal{S}\right),
\ee
where $L\left(\mathcal{S}\right)$ is the number of non-identity elements in the Pauli string $\mathcal{S}$. The operator size starts from the minimum value of unity when the operator $M_D$ is a single-qubit local operator, and saturates to the maximum value of $3n/4$ when it is uniformly distributed in the space spanned by Pauli strings. We numerically study the ensemble-averaged operator size with different VQC architectures in Fig.~\ref{fig:operator_size}, when each two-qubit gate is randomly chosen. Comparing the operator size growth with the performances in state discrimination and generation of Fig.~\ref{fig:Haar_architectures}, We find the same ordering for all VQC architectures. This consistency confirms the connection between the scrambling power and the discrimination power of VQCs.

\subsection{Performance of simplified gate sets}
\label{sec:nisq}

In the NISQ era~\cite{Preskill2018quantumcomputingin}, quantum circuit implementations are limited due to device imperfections. In particular, the imperfections accumulate with the increase of the number of quantum gates and the depth of the circuit. In all the VQC architectures being explored in Sec.~\ref{sec:architecture}, the realization of a universal two-qubit gate in fact requires three CNOT gates and additional qubit rotations (see Appendix~\ref{app:local_gates}), creating extra burdens in the VQC implementation. Another major constrain comes from the vanishing gradient due to barren plateau~\cite{cerezo2021cost} that prevents the efficient training of VQCs, which limits the scale of the implementations.
In this section, we consider different ways to simplify the quantum gates in the brickwall VQC and probe the induced change of the performance and trainability in state discrimination. 

\begin{table}[b]
\begin{center}
\begin{tabular}{|c|c|c|c|}
\hline
Number & brickwall & sVQC & real sVQC \\
\hline
Parameters& %$15\left[(n-1)\lfloor\frac{D}{2}\rfloor+\lfloor\frac{n}{2}\rfloor(D\mod{2})\right]$
$\sim \frac{15}{2}nD$ & $3n(D+1)$ & $n(D+1)$ \\
\hline
CNOT/CZ gates & %$3\left[(n-1)\lfloor\frac{D}{2}\rfloor+\lfloor\frac{n}{2}\rfloor(D\mod{2})\right]$
$\sim \frac{3}{2}nD$ & 
%$\lfloor\frac{n}{2}\rfloor D$ 
$\sim \frac{1}{2}nD$ & 
%$\lfloor\frac{n}{2}\rfloor D$
$ \sim \frac{1}{2}nD$ \\
\hline
\end{tabular}  
\caption{The number of parameters and CNOT/CZ gates of brickwall ansatz, sVQC and real sVQC ansatz to the leading order $O(nD)$ with depth $D$ in a system of $n$ qubits. 
\label{table:para}
}  
\end{center}
\end{table}

\begin{figure}[t]
    \centering
    \includegraphics[width=0.475\textwidth]{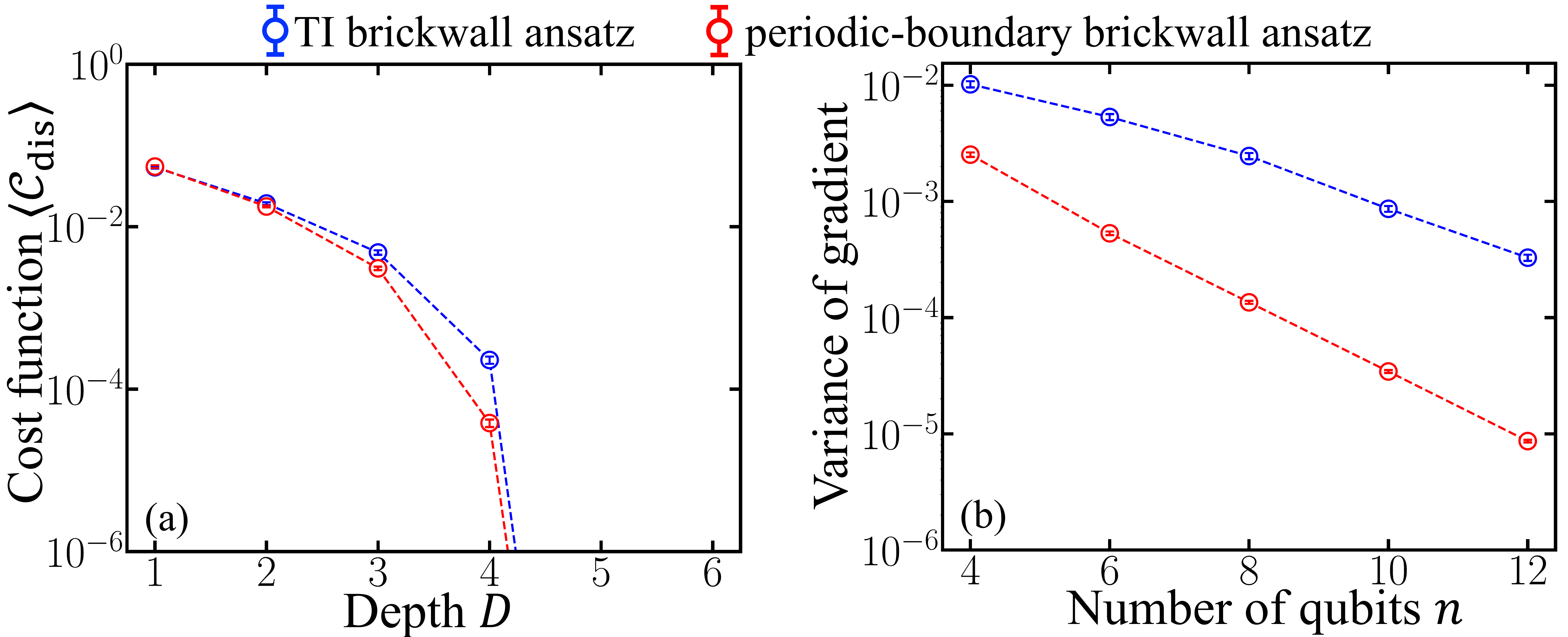}
    \caption{Cost function $\braket{\calC_{\rm dis}}$ (a) and average variance of gradient $\braket{{\rm Var}\left(g_i\right)}_i$ (b) of different brickwall ansatz in discriminating between random states sampled from $\mathbb{S}(2^n)$ with number of qubits $n=6$. We take TI and periodic-boundary brickwall ansatz. In (b) all ansatzs are set to be $D=2$.}
    \label{fig:tsSC}
\end{figure}

As we are often considering state discrimination between TI states, a natural attempt to simplify the VQC is to enforce TI on each layer of the VQC, including the gate parameters and a periodic boundary. As the TI symmetry reduces the number of parameters, we expect TI VQCs to be more trainable, which is confirmed by the gradient evaluations in Fig.~\ref{fig:tsSC}(b): the TI VQC typically shows a much larger gradient. Although assuming symmetry might lose some performance, however, as show in Fig.~\ref{fig:tsSC}(a), we find that when the input is symmetric, the TI brickwall ansatz (blue) provides almost identical performance to the periodic-boundary brickwall ansatz without the symmetry constraint (red). This shows that enforcing the VQC to have the same symmetry as the input simplifies the training of the VQC, while not losing much performance.

\begin{figure}
    \centering
    \includegraphics[width=0.475\textwidth]{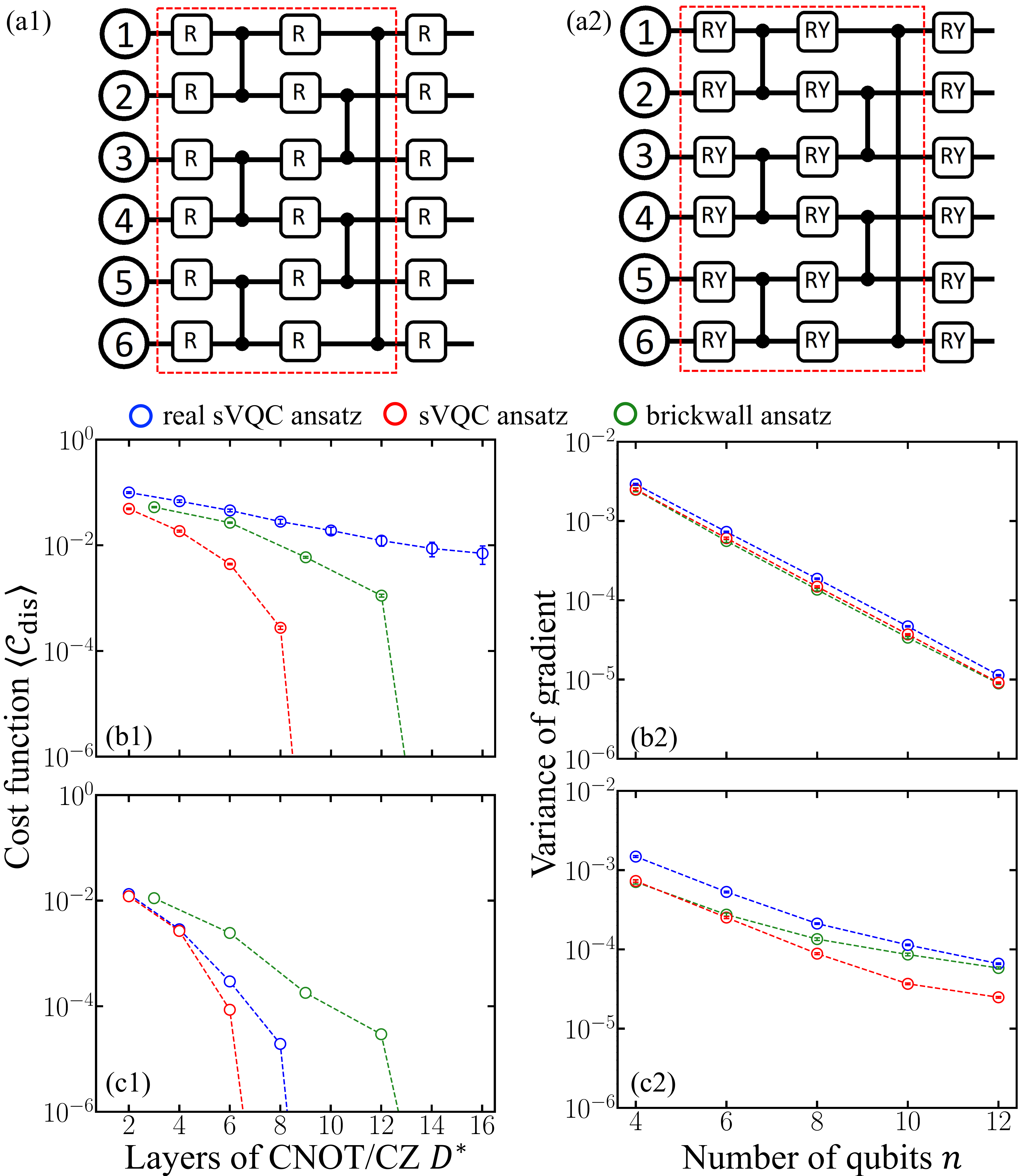}
    \caption{Architectures and performance for nisq VQCs. (a) Layout of an $n=6$-qubit sVQC ansatz and real sVQC ansatz, plotted in (a1) and (a2) separately. The sVQC ansatz consists of CZ gates and generic single qubit rotations, and the real sVQC ansatz consists of CZ gates and RY rotations. The circuits surrounded by the red dashed box represent $D^\star=2$ layers and at the end of circuit each qubit is applied with a rotation. (b1)-(b2) Cost function $\braket{\calC_{\rm dis}}$ and average variance of gradient $\braket{{\rm Var}\left(g_i\right)}_i$ for discriminating between Haar random states. (c1)-(c2) Cost function and average variance of gradient for discriminating between TFIM ground states with $g=1, 10$. We benchmark in a system of $n=6$ qubits. In (b2) and (c2) we take $D^\star = 6$ for all ansatz. Note for the brickwall ansatz, it corresponds to $D=2$.}
    \label{fig:nisq}
\end{figure}

Next, we consider simplifying the set of gates in VQCs. We replace each universal two-qubit gate in Fig.~\ref{fig:architectures}(a) (which requires three CNOT gates) with a single CZ gate \footnote{an extra CZ gate is inserted every two layers to form a periodic boundary condition}, and insert general single-qubit rotations in between each layer of CNOT gates, leading to the simple VQC (sVQC) of Fig.~\ref{fig:nisq}(a1) similar to the designs in Refs.~\cite{kandala2017hardware, havlivcek2019supervised}. In an sVQC, each single-qubit rotation $R(\theta_1, \theta_2, \theta_3) \equiv e^{-i\theta_1 Z/2}e^{-i\theta_2 Y/2}e^{-i\theta_3 Z/2}$ is charaterized by three angles $\theta_1,\theta_2$ and $\theta_3$, where $Z, Y$ are Pauli matrices. We can further reduce the complexity of each single-qubit gate by constraining its matrix representation to be real (via fixing $\theta_1=\theta_3=0$), leading to the real sVQC architecture which only implements real unitary $U_D$ in Fig.~\ref{fig:nisq}(a2). Real sVQCs are widely utilized in eigen-solvers~\cite{bravo2020scaling, wiersema2020exploring}, as the ground state of a time-reversal-symmetric Hamiltonian can be taken as real. The overall number of parameters (which equals the number of single qubit gates) and the number of CNOT/CZ gates are listed in Table~\ref{table:para} for comparison: both sVQC and real sVQC reduce the number of CNOT/CZ gates by a factor of three; while sVQC reduces the number of parameters roughly by half, the real sVQC reduces the number of parameters roughly by a factor of seven.

To ensure a fair comparison of error probability performance and trainability, instead of the depth $D$ of VQCs in terms of universal two-qubit gates that we consider in most of the paper, we count the number of layers of CNOT/CZ gates $D^\star$ in the final physical implementations in the original (brickwall) ansatz, sVQC ansatz and real sVQC ansatz. In Fig.~\ref{fig:nisq}, we find that sVQC outperforms the original ansatz consistently at the same $D^\star$, while the trainability barely changes. This is due to a certain level of redundancy in requiring each two-qubit gate to be universal in the original ansatz. The real sVQC further restricts the unitary implemented by the VQC to be real without losing much performance compared to sVQC in the discrimination between real ground states of TFIM, as shown in Fig.~\ref{fig:nisq}(c1); indeed, the trainability of the real sVQC improves due to the further simplification, as indicated by the larger gradients shown in Fig.~\ref{fig:nisq}(c2). While for Haar random states, due to the lack of symmetry of the input, such a restriction to real unitaries harms the performance while not improving the trainability, as shown in Fig.~\ref{fig:nisq}(b1)(b2).

\section{Discussions and Conclusions}
Adopting VQCs to the symmetry of the problem will maintain the discriminative power for the particular problem while simplifying the VQC implementations. For example, when the input is TI, constraining the VQC to be TI not only reduces the number of parameters and improves the trainability, but also preserves the performance; Similar advantages apply to assuming real VQCs for real ground states of TFIM. 
However, oversimplification can be problematic. As we have seen in Fig.~\ref{fig:tsSC}, constraining the unitary to be real will significantly harm the performance when discriminating between general complex quantum states.
As in the typical case, powerful VQCs are hard to train due to small gradients~\cite{cerezo2021cost}, one needs to utilize the structure and symmetry of the task to simplify the circuit and enable efficient training. Our results indicate that such a simplification needs to be tailored with caution.

To conclude, we propose an MLE-VQC scheme for quantum data classification, which shows an error probability advantage over the standard approach. As the depth of the VQC increases in an extensive way, the error probability of VQCs decreases exponentially towards the Helstrom limit. Despite being popular choices, non-extensive VQCs such as QCNN and MERA are sub-optimal in their error probability performances. 
The proposed MLE-VQC scheme can be implemented on near-term quantum device, and has the potential to be used in various applications. It is an important future direction to explore the MLE-VQC's performance in different applications, as in each application the symmetry and structure of the problem vary and may allow additional simplifications.

\begin{acknowledgments}
This material is based upon work supported by the U.S. Department of Energy, Office of Science, National Quantum Information Science Research Centers, Superconducting Quantum Materials and Systems Center (SQMS) under the contract No. DE-AC02-07CH11359. QZ acknowledges support from the Defense Advanced Research Projects Agency (DARPA) under Young Faculty Award (YFA) Grant No. N660012014029.
\end{acknowledgments}

\appendix

\section{Preliminary of state discrimination } 
\label{app:preliminary}

\subsection{General Helstrom limit}

The minimum `Helstrom' error probability~\cite{Helstrom_1967,Helstrom_1976} for the discrimination between $m$ states $\{\rho_i\}_{i=0}^{m-1}$ with prior probability $\{p_i\}_{i=0}^{m-1}$ 
\be
P_{\rm H}\left(\{\rho_i, p_i\}\right)=1-\max_{\sum_i \Pi_i=I}\sum_i p_i {\rm Tr}\left(\rho_i \Pi_i\right),
\label{Helstrom_general}
\ee
where the POVM element $\Pi_i$ corresponds to the hypothesis that the state is $\rho_i$. For the binary pure-state case, Eq.~\eqref{Helstrom_general} can be reduced to Eq.~\eqref{Helstrom_pure}.

\subsection{MLE decision for general state discrimination}
Consider the general state discrimination described above, when the VQC implements the unitary $U_D$ on input $\rho_i$, the measurement result $j$ appears with probability
\be 
P(j|\rho_i) = \Tr{\ket{j}\bra{j} U_D\rho_i U_D^\dagger},
\ee 
where $\{j\}_{j=0}^{2^n-1}$ forms the set of all possible measurement results. Conditioned on the measurement result $j$, MLE strategy minimizes the average error probability by making the decision on the state $\rho_{\tilde{i}}$ via
\be 
\tilde{i}(j)=\argmax_i p_i P(j|\rho_i),
\ee 
leading to the minimum error probability for a fixed measurement choice as
\be
P_{\rm E}\left(\{\rho_i,p_i\}\right) = 1-\sum_{i=0}^{m-1} p_i P(\tilde{i}=i|\rho_i).
\ee
Here the conditional correct probability
$
P(\tilde{i}=i|\rho_i) = \sum_{\tilde{i}(j)=i} P(j|\rho_i).
$

\begin{figure}[t]
    \centering
    \includegraphics[width=0.35\textwidth]{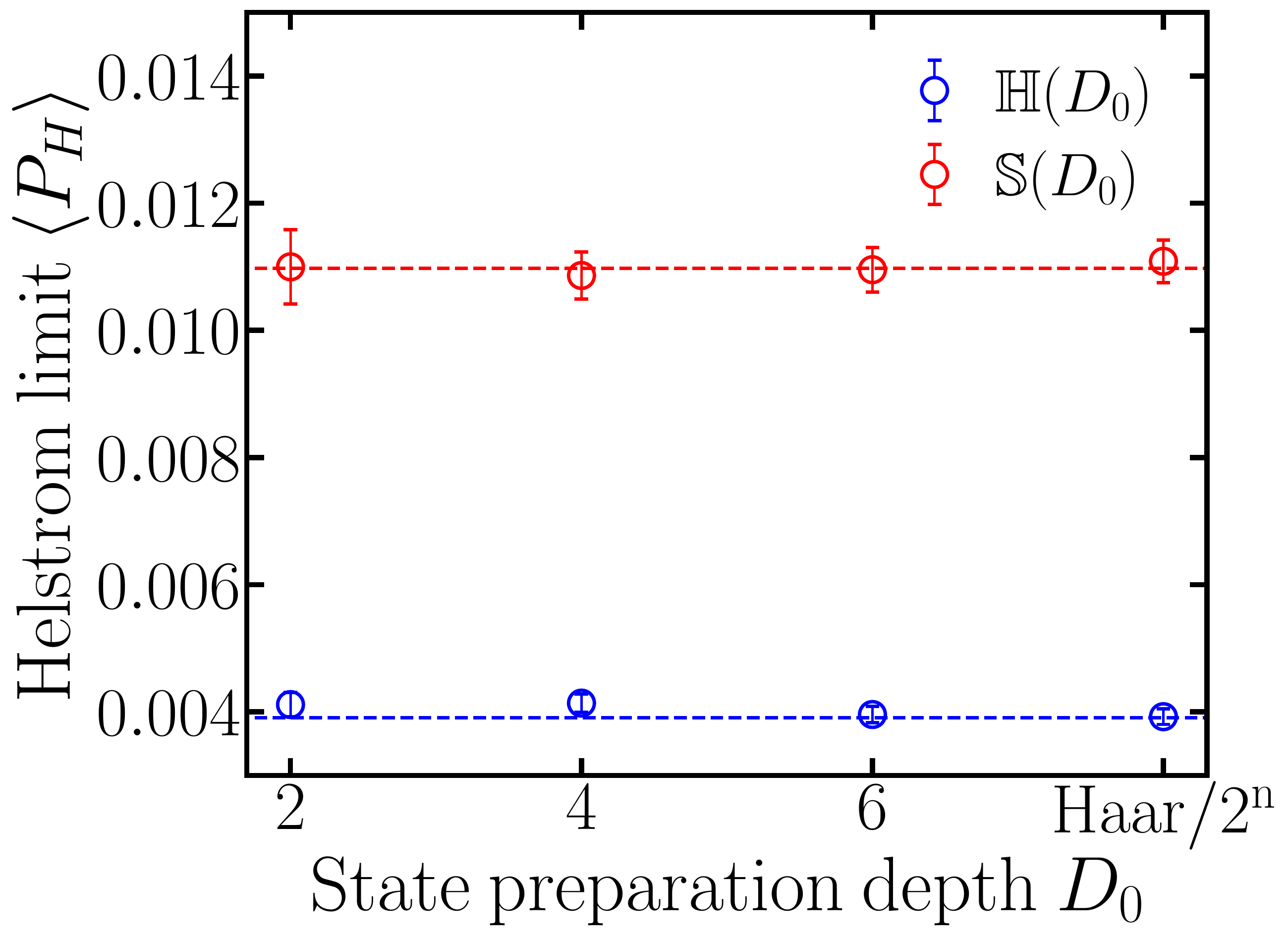}
    \caption{Ensemble-averaged Helstrom limit $\braket{P_H}$ for $n=6$ qubits states sampled from $\mathbb{H}(D_0)$ and $\mathbb{S}(D_0)$. The blue dashed line represents $\braket{P_H} \sim 1/2^{n+2}$ and the red dashed line is the average over red dots.}
    \label{fig:mean_helstrom}
\end{figure}

In this paper, we focus on the binary pure state case with equal prior and the MLE error probability is reduced to
\be
P_{\rm E} =  \frac{1}{2}\left\{1-\sum_{\substack{j:
P(j|\psi_0)\ge P(j|\psi_1)}} \left[P\left(j|\psi_0\right)- P\left(j|\psi_1\right)\right]\right\},
\label{PE}
\ee
where $P(j|\psi_i)$ is probability of deciding the state is $\psi_j$ given the true input $\psi_i$.

\section{Evaluation of $\braket{P_{\rm H}\left(\psi_0,\psi_1\right)}_{\mathbb{H}\left(D_0\right)}$}
\label{app:haar}
\subsection{$n\gg1$ limit of finite $D_0$}

When $n\gg1$, we expect the typical state overlap $|\braket{\psi_0|\psi_1}|^2$ to be small, therefore
\be 
P_{\rm H}\left(\psi_0,\psi_1\right)=\frac{1}{2}\left[1-\sqrt{1-|\braket{\psi_0|\psi_1}|^2}\right]\sim \frac{1}{4}|\braket{\psi_0|\psi_1}|^2.
\ee 
Below we show that regardless of $D_0$, within the state ensemble $\mathbb{H}\left(D_0\right)$, the typical Helstrom limit between states is $\braket{P_{\rm H}\left(\psi_0,\psi_1\right)}_{\mathbb{H}\left(D_0\right)}\sim 1/2^{n+2}$, which simply follows from the typical overlap $\braket{|\braket{\psi_0|\psi_1}|^2}_{\mathbb{H}\left(D_0\right)}=1/2^n$. We also show it numerically in Fig.~\ref{fig:mean_helstrom}.

\begin{figure}[t]
    \centering
    \includegraphics[width=0.4\textwidth]{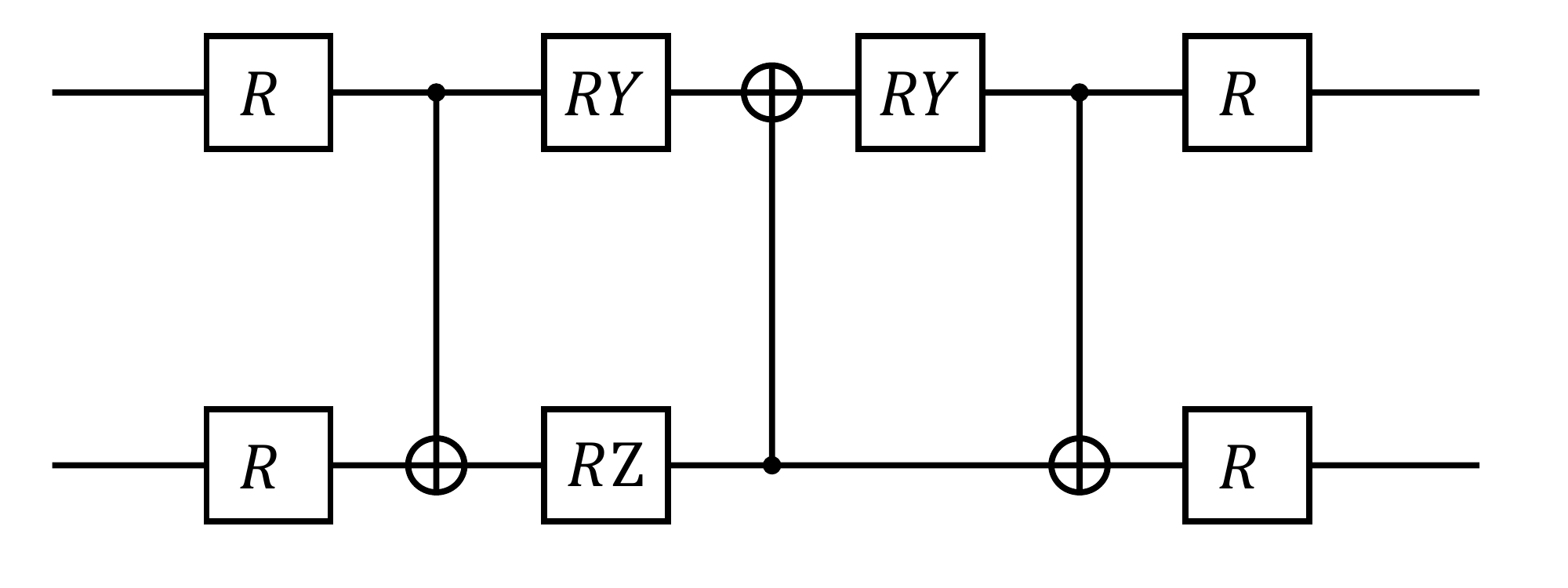}
    \caption{A decomposition of a general two-qubit gate. The $R$ gate represents an arbitrary single qubit rotation with $3$ independent parameters as $R(\theta_1, \theta_2, \theta_3)=RZ(\theta_1)RY(\theta_2)RZ(\theta_3)$.}
    \label{fig:U2_gate}
\end{figure}

\begin{figure}[b]
    \centering
    \includegraphics[width=0.475\textwidth]{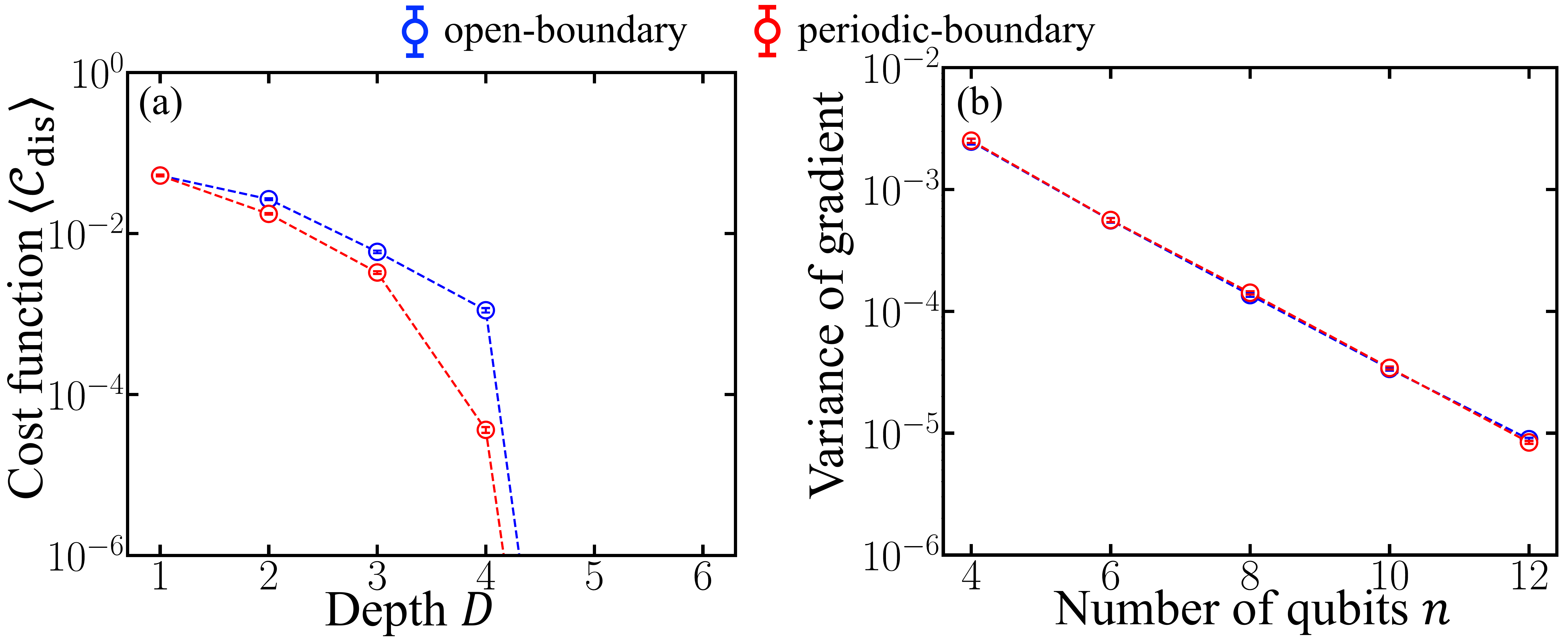}
    \caption{(a) Cost function $\braket{\calC_{\rm dis}}$ and (b) average variance of gradient $\braket{{\rm Var}\left(g_i\right)}_i$ for discriminating between $n=6$ qubits Haar random states using open-boundary and periodic-boundary brickwall ansatzs. In (b) we evaluate both ansatzs at the depth of $D=2$.}
    \label{fig:boundary}
\end{figure}

\begin{figure*}
    \centering
    \includegraphics[width=0.7\textwidth]{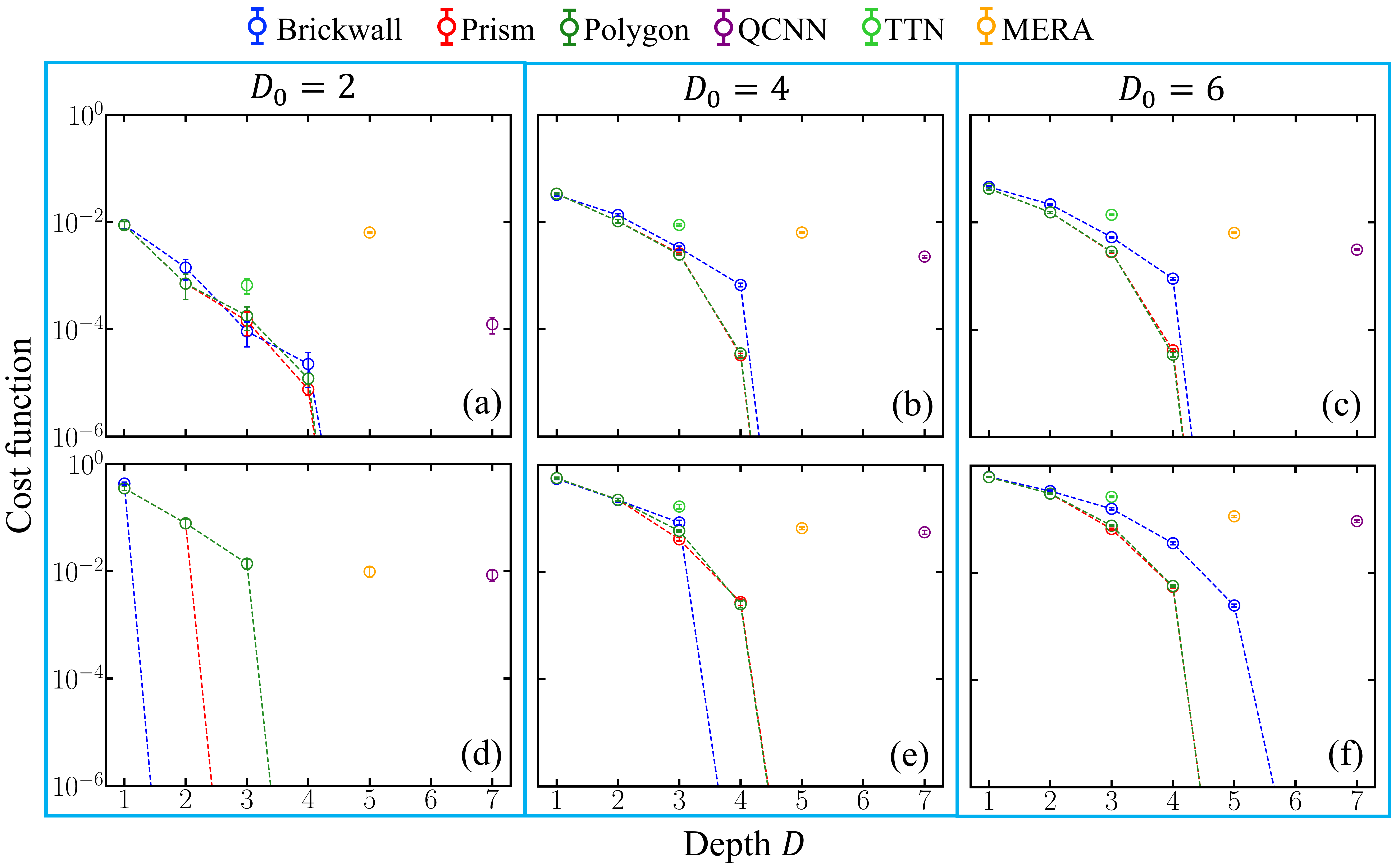}
    \caption{Cost functions of different VQC architectures in the discrimination (top) and generation (bottom) of random states sampled from $\mathbb{H}(D_0)$ with $D_0=2, 4, 6$ in a system of $n=6$ qubits. 
    \label{fig:SCob_archs}
    }
    \centering
    \includegraphics[width=0.7\textwidth]{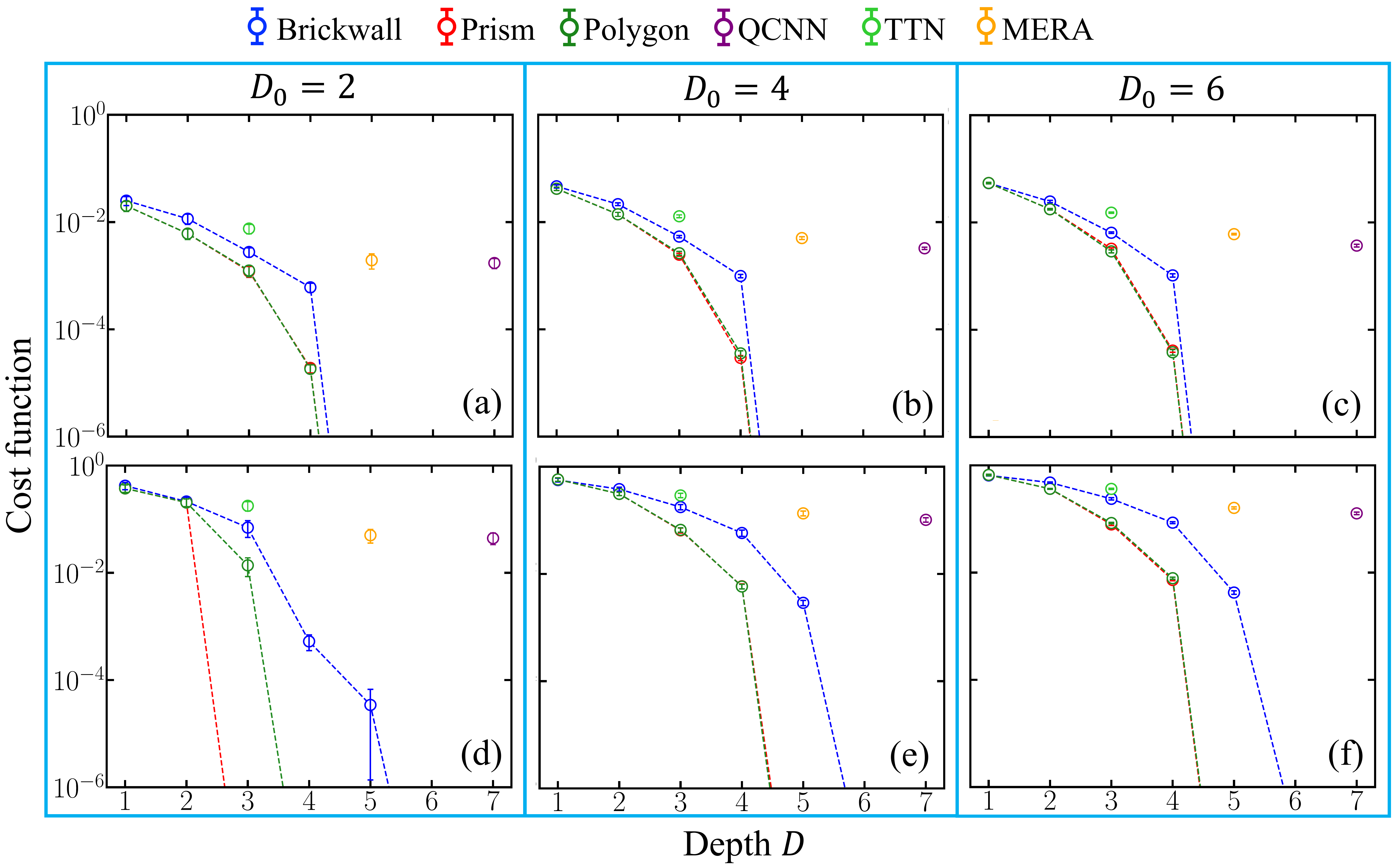}
    \caption{Cost functions of different VQC architectures in discrimination (top) and generation (bottom) of TI random states sampled from $\mathbb{S}(D_0)$ with $D_0=2, 4, 6$ in a system of $n=6$ qubits. 
    \label{fig:tsSC_archs}
    }
\end{figure*}

Consider an $n$-qubit system, suppose one considers an initial product state $\ket{0}^{\otimes n}$, for two local quantum circuits implementing unitaries $U_0,U_1$, the overlap
\be 
c(U_0,U_1)=\braket{\bm 0 |U^\dagger|\bm 0}\braket{\bm 0 |U|\bm 0},
\ee 
where $U=U_0^\dagger U_1$. As $U$ and $U^\dagger$ each appears only once, taking the ensemble average over a 1-design~\cite{gross2007evenly,ambainis2007quantum,roberts2017chaos} suffices to produce the Haar value.
In our case, we consider random local quantum circuits for $U_0$ and $ U_1$, with each two-qubit unitary Haar random. Regardless of the number of layers of the circuit, the ensemble $\{U_0^\dagger U_1\}$ forms a 1-design, therefore
\be
\overline{c(U_0,U_1)}
=\int_{\rm Haar} dU\braket{\bm 0 |U^\dagger|\bm 0}\braket{\bm 0 |U|\bm 0}=\frac{1}{2^n},
\ee  
where we utilized the Haar average identity for two elements of a $U(d)$ matrix,
\be 
\int_{\rm Haar} dU U_{\alpha a}U^*_{\beta b} = \frac{1}{d}\delta_{\alpha\beta}\delta_{ab}.
\ee 
where $d$ is its dimension.

\subsection{$D_0\gg1$ limit}
The exact $\braket{P_{\rm H}\left(\psi_0,\psi_1\right)}_{\mathbb{H}\left(D_0\right)}$ is not easy to compute for a finite $D_0$. Here we consider the case of $D_0\gg1$, where the ensemble $\mathbb{H}\left(D_0\right)$ is simply Haar random. In this case, the distribution of the overlap $x=|\braket{\psi_0|\psi_1}|^2$ can be analytically obtained~\cite{zhuang2013equilibration}.

Consider $d=2^n$ complex numbers $\{\alpha_i\}_{i=1}^d$ as the complex amplitudes of the the states $\psi_1$ with normalization condition $\sum_{i=1}^d |\alpha_i|^2 = 1$, it indicates that the amplitudes $\{\alpha_i\}_{i=1}^d$ forms $2d$-sphere. The other state $\psi_0$ we can choose it to be $\ket{\psi_0} = \ket{\mathbf{0}}$ with the freedom in choosing the Haar unitary. The probability distribution for $x=|\braket{\psi_0|\psi_1}|^2 = |\braket{\bf{0}|\psi_1}|^2 = |\alpha_1|^2=\gamma$ is
\begin{equation}
\begin{split}
    P(x = \gamma) & = \frac{\int\prod_id^2 a_i \delta\left(\gamma-|\alpha_1|^2\right)\delta\left(1-\sum_i |\alpha_i|^2\right)}{\int\prod_i d^2 a_i \delta\left(1-\sum_i |a_i|^2\right)}\\
    & = \frac{\int\prod_{i>1} d^2 a_i S_1\left(\sqrt{\gamma}\right) \delta\left(1-\gamma-\sum_{i>1}|\alpha_i|^2\right)}{2\sqrt{\gamma}S_{2d-1}(1)/2}\\
    & = \frac{S_1\left(\sqrt{\gamma}\right)S_{2d-3}\left(\sqrt{1-\gamma}\right)}{2\sqrt{\gamma}\sqrt{1-\gamma}S_{2d-1}(1)}\\
    & = (d-1)(1-\gamma)^{d-2}.
\end{split}
\end{equation}
Here we utilize the $N$-dimensional sphere surface area formula
\begin{equation}
\begin{split}
    S_N(R) & \equiv \int \prod_{l=1}^{N+1} dx_l \delta\left(R-\sqrt{\sum_{l=1}^{N+1} x_l^2}\right)\\
    & = 2R\int \prod_{l=1}^{N+1} dx_l \delta\left(R^2-\sum_{l=1}^{N+1} x_l^2\right) = \frac{2\pi^{\frac{N+1}{2}}}{\Gamma\left(\frac{N+1}{2}\right)}R^N,
\end{split}
\end{equation}
where we use the expansion $2a\delta(x^2-a^2) = \delta(x-a)+\delta(x+a)$. One can also check that the probability $\int_0^1 d\gamma P(x=\gamma)=1$ is normalized.
From the probability distribution, we have the Haar average Helstrom limit as
\begin{equation}
\begin{split}
    \expval{P_{\rm H}}_{\rm Haar} & = \int_0^1 d\gamma \frac{1}{2}\left(1-\sqrt{1-\gamma}\right)p(x=\gamma)\\
    & = \frac{d-1}{2\left(d^2-3d+1\right)}\\
    & = \frac{1}{2\left(2^{n+1}-1\right)}.
\end{split}
\end{equation}
When $n \gg 1$, the Haar average Helstrom limit $\expval{P_{\rm H}}_{\rm Haar} \sim 1/2^{n+2}$.

\section{Local random gates construction}
\label{app:local_gates}

As shown in Fig.~\ref{fig:architectures}, various architectures of VQC are constructed from local two-qubit gates. In general, a two-qubit gate, in the form of $4\times 4$ unitary, includes $15$ independent parameters (up to a global phase). Such a gate can be decomposed into single qubit rotations and up to $3$ CNOT gate~\cite{vatan2004optimal}, as we show in Fig.~\ref{fig:U2_gate}.

\section{Boundary condition for VQCs}
\label{app:boundary}

In Sec.~\ref{sec:local_VQC}, we assume the brickwall VQC ansatz to be open-boundary, which is a local quantum circuit. However, one can also assume the brickwall ansatz to have a periodic-boundary. In Fig.~\ref{fig:boundary}, we compare the performance and gradient of open-boundary and periodic-boundary brickwall ansatzs in discriminating between Haar random states. Just as we expect, the periodic-boundary brickwall ansatz performs slightly better than the open-boundary one in the low depth region, due to the extra gates on the first and last qubit that help process global quantum information, while the gradient is nearly the same which leads to the same difficulty of training. Therefore, without specific clarification we utilize open-boundary brickwall ansatz in the main text to restrict the locality of VQCs.

\begin{figure}[t]
    \centering
    \includegraphics[width=0.475\textwidth]{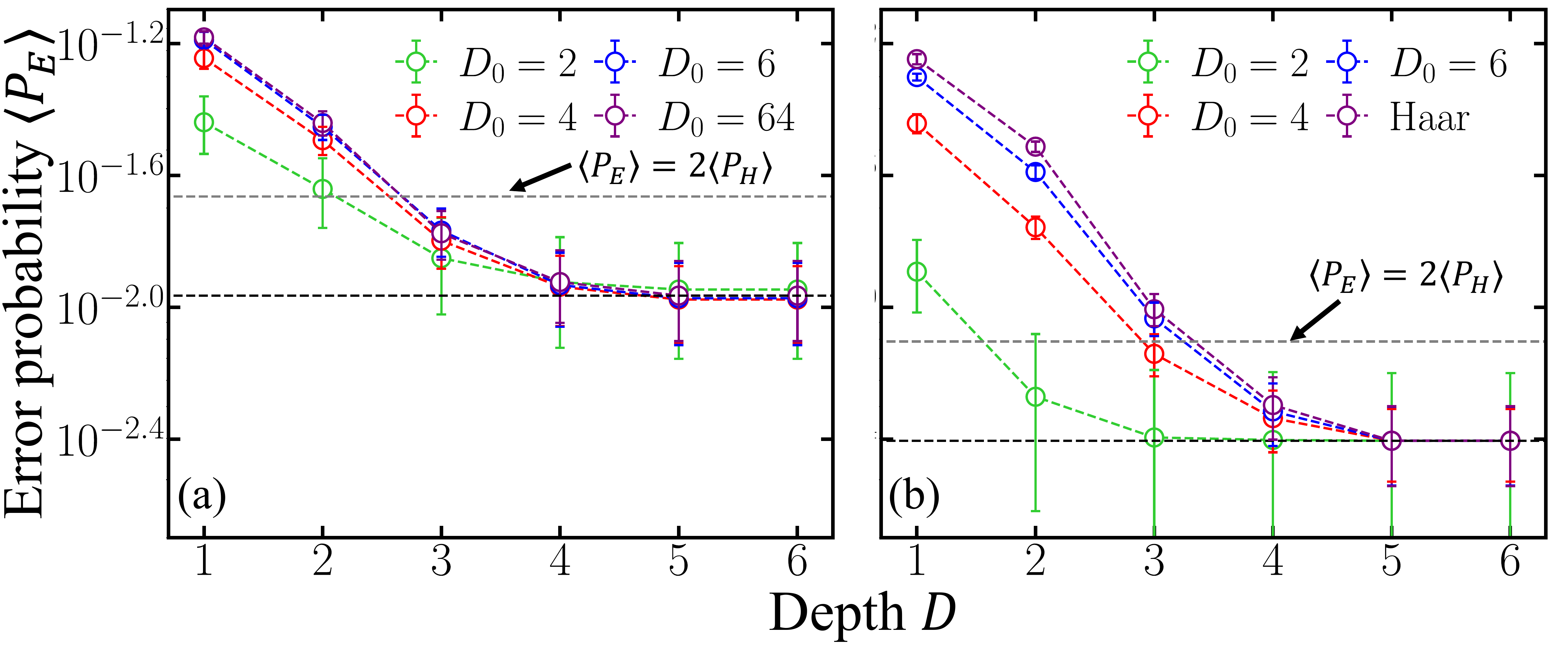}
    \caption{Error probability for discriminating between random states from $\mathbb{S}(D_0)$ (left) and $\mathbb{H}(D_0)$(right) using brickwall ansatz in an $n=6$ system. Black and grey dashed lines in (a) and (b) show ensemble-averaged Helstrom limit $\braket{P_H}$ and $2\braket{P_H}$.}
    \label{fig:symmetries}
    \centering
    \includegraphics[width=0.475\textwidth]{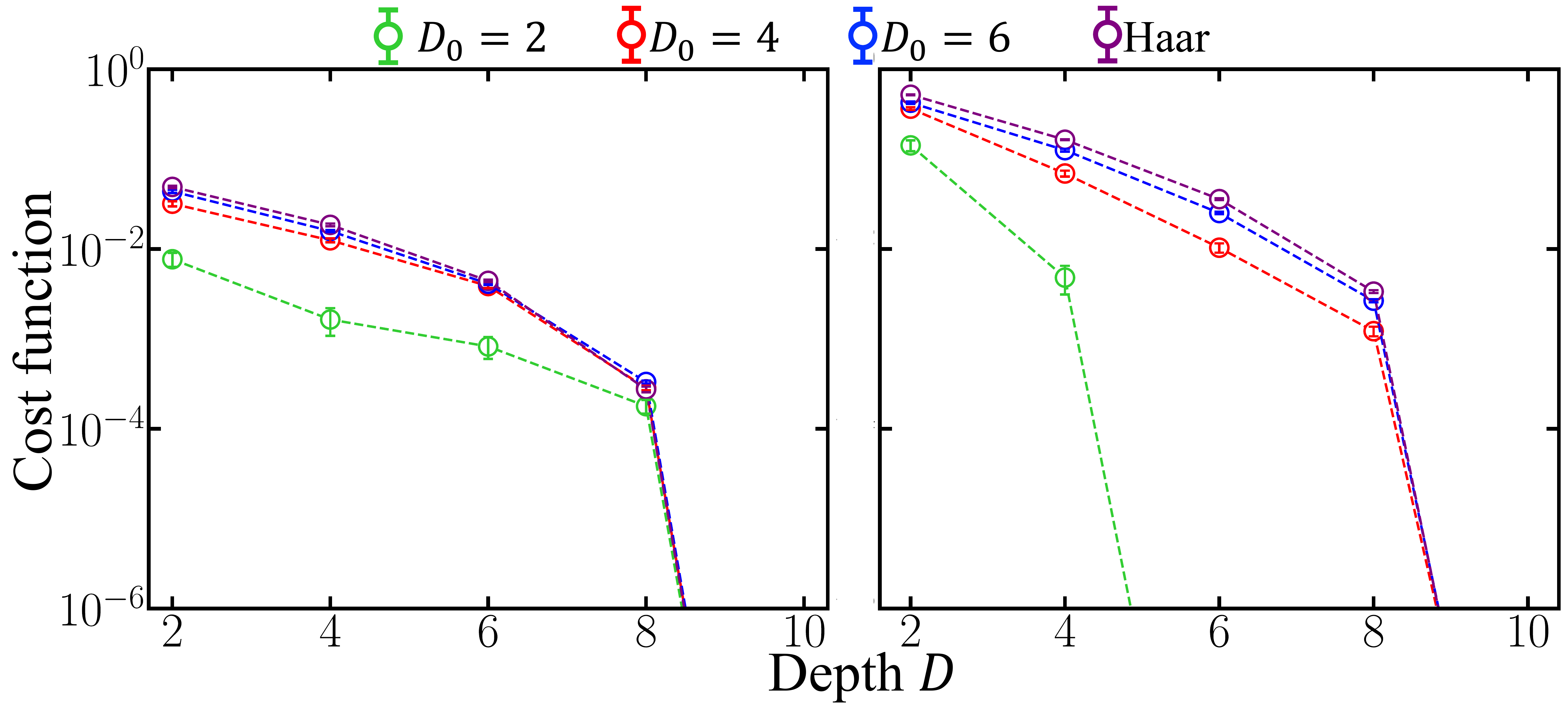}
    \caption{Cost functions of the sVQC ansatz in discrimination $\braket{\calC_{\rm dis}}$ (a) and generating $\braket{\calC_{\rm gen}}$ (b) of random states from $\mathbb{H}(D_0)$ with $D_0=2, 4, 6$ and Haar random states in a system of $n=6$ qubits. In this case, the depth $D=D^\star$ equals the layers of CNOT/CZ.}
    \label{fig:sVQC_all}
\end{figure}
\section{More results of VQCs performance}
\label{app:other_performance}

The VQC architectures utilized in Sec.~\ref{sec:benchmarks} are applied to the discrimination and generation of random states sampled from $\mathbb{H}(D_0)$ and $\mathbb{S}(D_0)$ with $D_0 \le n$. With $D_0$ increases, the random states become more complex and entangled, as shown in Fig.~\ref{fig:state_prepare}(b). Similar to the Haar case, we can also confirm the advantage for extensive architectures towards non-extensive ones in small $D_0$ cases from Fig.~\ref{fig:SCob_archs}. From Fig.~\ref{fig:SCob_archs}(a)(b)(c), as states become more complex, for a fixed $D$ the residual error also increases with extensive architectures, brickwall, prism and polygon. Compared to generation tasks shown in Fig.~\ref{fig:SCob_archs}(d)(e)(f), discrimination still shows better performance within the depth range that $\calC_{\rm gen}(U_D, \ket{\psi})$ is non-zero.

Unlike the Haar case shown in Fig.~\ref{fig:Haar_architectures}, we can find that as $D_0$ decreases, the advantage of prism and polygon ansatzs over brickwall ansatz is reduced. Note that the information of input states is locally stored as $D_0$ is small, and thus the non-local gates in those architectures (which help to processes global information) could not perform as efficiently as in the large $D_0$ cases. The above discussions can also be extended to the performance of VQCs in TI states from $\mathbb{S}(D_0)$, as we show in Fig.~\ref{fig:tsSC_archs}.

In Fig.~\ref{fig:SC_gap}(a), we see that the discrimination residual error for random states from Haar ensemble and $\mathbb{S}(2^n)$ roughly agrees when $D$ is low. We plot the discrimination error probability directly in Fig.~\ref{fig:symmetries}. Compared to the typical random states, the Helstrom limit of TI random states is larger and independent of $D_0$, as we already confirmed in Fig.~\ref{fig:mean_helstrom}. Except for that, for all $D_0$, the error probability for TI states is still exponential suppressed with the depth $D$.

Finally, we provide the performance of the sVQC anstaz, in Fig.~\ref{fig:sVQC_all}. Due to the limited number of CNOT/CZ gates and parameters, the depth required to approach the ultimate limit increases for the sVQC ansatz.

\section{Computation details}
\label{app:comp_details}
\begin{figure}
    \centering
    \includegraphics[width=0.3\textwidth]{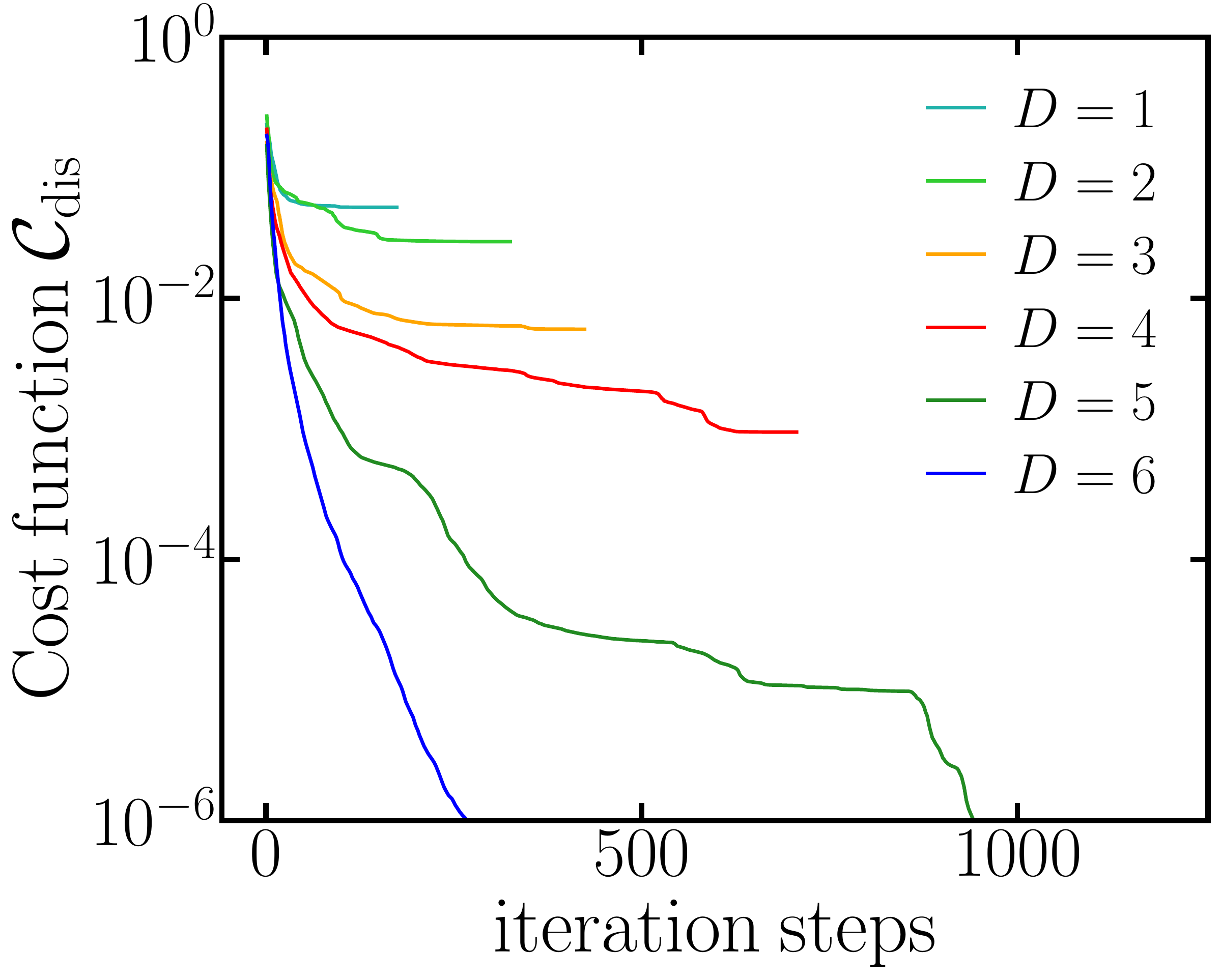}
    \caption{The optimization history of cost function $\calC_{\rm dis}$ in discriminating between a pair of Haar random states. We apply the brickwall ansatz with different depth $D$ in a system of $n=6$ qubits.}
    \label{fig:train_hist}
\end{figure}

To implement different VQC ansatzs, we use Qulacs~\cite{suzuki2020qulacs}, a high performance VQC simulator for Python. We utilize the BFGS algorithm~\cite{virtanen2020scipy}, a gradient based quasi-Newton method for numerical optimization on parameters of VQCs.
Compared to regular gradient descent training methods like stochastic gradient descent (SGD) and Adam used in classical neural networks, the VQC can be optimized in less iteration steps using the BFGS algorithm. We choose the best one among the $40$ optimization results from random initializations of parameters. An example of optimization history is shown in Fig.~\ref{fig:train_hist}.

We evaluate the gradient of cost functions with respect to each parameter using the central finite difference method. Another formalism, parameter-shift rule, is also proposed to provide an analytical expression for the gradient~\cite{mitarai2018quantum}. For a given observable $O$, the gradient of the mean of the observable is $g_{\rm PSR}^i\equiv \partial_i \braket{O}=\frac{1}{2}\left(\braket{O}^+ - \braket{O}^-\right)$, where $\braket{O}^{\pm} = \braket{O}_{\theta_i \pm \frac{\pi}{2}}$. Note that in the case of discrimination with MLE, the classical post-processing highly depends on the initial parameters, and thus only the finite difference is available for the gradient calculation. For a fair gradient comparison in all cases, we adopt finite-difference in calculations of gradient. The step size $\Delta s$ is also a free parameter in the definition of finite difference, and here we apply $\Delta s=10^{-6}$.

%\bibliography{myref.bib}

\end{document}